\documentclass[a4paper]{jpconf}

\usepackage{lmodern}
\usepackage{amsmath}
\usepackage{amssymb}
\usepackage{graphicx}
\usepackage{nomencl}
\usepackage{cite}
\usepackage{esint}
\usepackage{geometry}
\usepackage{float}
\usepackage{hhline}

\begin{document}

\title{Symmetric Galerkin boundary element method for computing the quantum states of the electron in a piecewise-uniform mesoscopic system}%

\author{Andrea Cagliero$^1$ and Lyes Rahmouni$^{1,2}$}

\address{$^1$At the time of the study, both authors were affiliated with the Microwaves Department of IMT Atlantique, Institut Mines-T\'{e}l\'{e}com, and with the Laboratory for Science and Technologies of Information, Communication and Knowledge Lab-STICC (CNRS), Brest, F-29238, France.}
\address{$^2$Department of Electronics and Telecommunications, Politecnico di Torino, I-10129 Torino, Italy.}

\ead{andrea.cagliero@edu.unito.it}

\begin{abstract}
The quantum behavior of charge carriers in semiconductor structures
is often described in terms of the effective mass Schr\"{o}dinger equation,
neglecting the rapid fluctuations of the wave function on the scale
of the atomic lattice. For systems with piecewise-constant mass and potential energy, this amounts
to solving a set of Helmholtz equations with wavenumbers dictated
by the physical parameters of each homogeneous subregion. Making use of the Green
function method, the system of differential equations can be expressed
in boundary integral form to enable efficient numerical solution.
In the present study, this strategy is applied in combination with
a Galerkin technique to compute the energy spectrum and the wave functions
of the electron in a mesoscopic structure composed of two regions.
The proposed formulation differs from those presented before for the
same scenario in that it implements a symmetric discretization of
the four Helmholtz boundary integral operators, which leads to compact
expressions and very accurate results.
\end{abstract}


\section{Introduction}

In recent years, fundamental and applied research in semiconductor
and solid state physics has undergone a significant evolution, with
particular emphasis on the study and development of mesoscopic structures
and quantum wells \cite{Balkanski2000,Harrison2016,Boer2018}. Indeed,
the quantum confinement of charge carriers is responsible for a rich
variety of phenomena that are of interest in optoelectronics, nanotechnology
and quantum computing \cite{Kulik2000}. As it is well-known, the
single-particle electronic properties of mesoscopic structures are
dictated by the Schr\"{o}dinger equation and depend on both the electron
energy and the confining potential of the atomic lattice. Since for
most geometries the energy levels and wave functions of the electron
inside such structures cannot be determined analytically, numerical
methods are required for both the analysis and interpretation of the
experimental results. Among the various computational techniques,
the finite element method (FEM) and the boundary element method (BEM)
have been explored in the literature \cite{Knipp1996,Gelbard2001,Ramdas2002,Cai2013,Sun2015,Klaseboer2017}.
Whereas the FEM consists in a volume discretization of the original
boundary value problem, the BEM leverages the Green function approach
to cast the partial differential equations into a boundary integral
form which is then projected on finite dimensional trial spaces. In
particular, this last step can be addressed by collocation, explicitly
imposing the boundary integral equations at a finite set of points,
or by the Galerkin approach, where the equations are enforced in a
weighted average sense \cite{Sutradhar2008}. The main limitation of the BEM is that it requires the knowledge of the Green function of the physical system: for arbitrary confining potentials, the Green function cannot be expressed in closed form. However, mesoscopic structures are often constitued by piecewise-homogeneous regions, which means that only free-space Green functions are needed. In this case, owing to the reduction in dimensionality, the BEM is often more efficient than the FEM; moreover,
the BEM provides a very natural strategy to compute the scattering
of the electron wave function into unbounded regions. Although most
of the literature deals primarily with collocation methods, the Galerkin
BEM is known to be more accurate and robust. 

In this work, a BEM involving the Galerkin discretization of the matrix
integral operator (\ref{eq:operatorH}) is proposed to solve the effective
mass Schr\"{o}dinger equation (Section \ref{sec:Schrodinger}) for a charge
carrier in a mesoscopic system comprising two regions with piecewise-constant mass and potential energy. The formulation
enables the determination of the discrete energy levels (Section \ref{sec:BoundStates}),
the scattering amplitudes (Section \ref{sec:Scattering}) and the
spectral density of the system (Section \ref{sec:Spectral}), as well
as the corresponding wave functions. For the reader's benefit, the paper 
is self-contained and in \ref{sec:A}, \ref{sec:B} and \ref{sec:C} some 
technical details of the BEM derivation are reminded to the reader and adapted to the notation of the main text.
Additional results and insights can be found in \ref{sec:D} and in \ref{sec:E}.

\section{Problem statement\label{sec:Schrodinger}}

Let us consider the non-relativistic time-dependent Schr\"{o}dinger equation
for the electron wave function within an arbitrary mesoscopic structure
\cite{Harrison2016}. Under the effective mass approximation for the
envelope function $\Psi\left(\mathbf{r},t\right)$, the equation
reads:
\begin{equation}
i\hbar\frac{\partial\Psi\left(\mathbf{r},t\right)}{\partial t}=-\frac{\hbar^{2}}{2}\nabla\cdot\left[\frac{1}{m\left(\mathbf{r}\right)}\nabla\Psi\left(\mathbf{r},t\right)\right]+V\left(\mathbf{r}\right)\Psi\left(\mathbf{r},t\right),\label{eq:timeDepSchrodinger}
\end{equation}
being $m\left(\mathbf{r}\right)$ the electron mass
\cite{LevyLeblond1995} and $V\left(\mathbf{r}\right)$ the
potential energy of the confining structure. Assuming a time-harmonic
dependence of the form:
\begin{equation}
\Psi\left(\mathbf{r},t\right)=\psi\left(\mathbf{r}\right)\exp\left(-i\frac{E}{\hbar}t\right),
\end{equation}
where $E$ represents the electron energy, equation (\ref{eq:timeDepSchrodinger}) is reduced to: 
\begin{equation}
\frac{\hbar^{2}}{2}\nabla\cdot\left[\frac{1}{m\left(\mathbf{r}\right)}\nabla\psi\left(\mathbf{r}\right)\right]+\left[E-V\left(\mathbf{r}\right)\right]\psi\left(\mathbf{r}\right)=0.\label{eq:timeIndepSchrodinger}
\end{equation}
For piecewise constant mass and potential energy, the previous expression
can be put in the same form as the scalar Helmholtz equation \cite{Knipp1996,Gelbard2001,Ramdas2002,Cai2013}.
Suppose, for instance, that the mesoscopic structure can be divided
into $N$ homogeneous subregions $\Omega_{j}$ such that $m\left(\mathbf{r}\right)=m_{j}$
and $V\left(\mathbf{r}\right)=V_{j}$ for $\mathbf{r}\in\Omega_{j}$,
where $j$ runs from $1$ to $N$. Within the $j$-th subregion, equation
(\ref{eq:timeIndepSchrodinger}) becomes:
\begin{equation}
\frac{\hbar^{2}}{2m_{j}}\Delta\psi\left(\mathbf{r}\right)+\left(E-V_{j}\right)\psi\left(\mathbf{r}\right)=0\quad\mathbf{r}\in\Omega_{j}.
\end{equation}
Here and below, the symbol $\Delta$ stands for the Laplacian. Making use of the following definition: 
\begin{equation}
k_{j}^{2}\equiv\frac{2m_{j}}{\hbar^{2}}\left(E-V_{j}\right),
\end{equation}
we finally obtain:
\begin{equation}
\Delta\psi\left(\mathbf{r}\right)+k_{j}^{2}\psi\left(\mathbf{r}\right)=0\quad\mathbf{r}\in\Omega_{j}.\label{eq:scalarHelmholtz}
\end{equation}
For each wavenumber $k_{j}$, the unique form of the free-space Green function $g_j\left(\mathbf{r},\mathbf{r}'\right)$ satisfying:
\begin{equation}
\Delta g_j\left(\mathbf{r},\mathbf{r}'\right)+k_{j}^{2}g_j\left(\mathbf{r},\mathbf{r}'\right)=-\delta\left(\mathbf{r}-\mathbf{r}'\right)
\end{equation}
and the Sommerfeld radiation condition:
\begin{equation}
\left| \frac{\mathbf{r}}{\left|\mathbf{r}\right|}\cdot \nabla g_j\left(\mathbf{r},\mathbf{r}'\right) -i k_{j} g_j\left(\mathbf{r},\mathbf{r}'\right) \right|
= O\left(\frac{1}{\left|\mathbf{r}\right|^2}\right)\quad \left|\mathbf{r}\right|\rightarrow \infty \label{eq:sommerfeld}
\end{equation}
will be considered, according to the usual convention.
It is important to note that, for the kinetic energy operator to be
Hermitian, both the wave function $\psi$ and its weighted normal
derivative $m^{-1}\partial_{n}\psi$ must be continuous across the
interface between any two different subregions \cite{LevyLeblond1995}.
In the next sections, these boundary conditions are used together with (\ref{eq:sommerfeld}) to solve
equation (\ref{eq:scalarHelmholtz}) numerically by the symmetric
Galerkin BEM for $N=2$ homogeneous subregions. 

\section{Bound states\label{sec:BoundStates}}

\subsection{Integral equations\label{subsec:IntegralEquations}}

\begin{figure}
\begin{centering}
\includegraphics[scale=0.3]{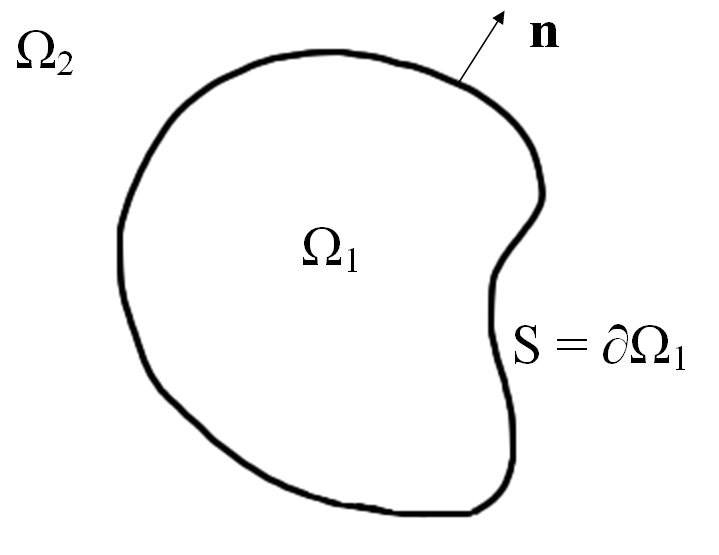}
\includegraphics[scale=0.3]{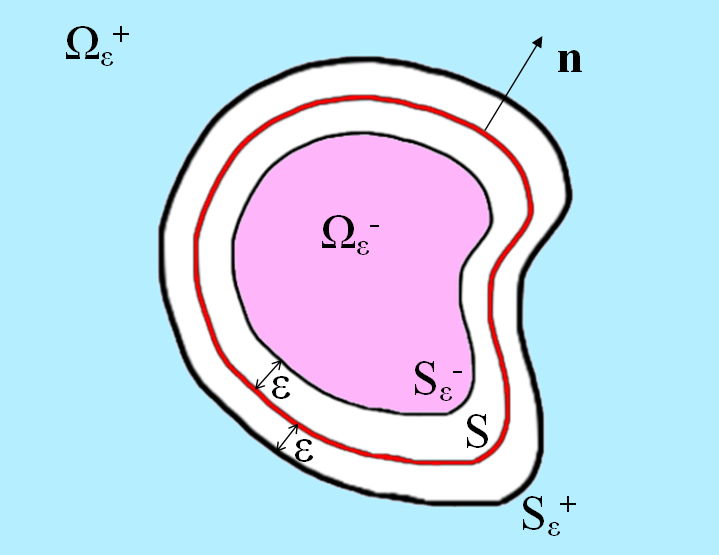}
\par\end{centering}
\caption{\textit{Left}: sketch of the geometry of the problem. \textit{Right}:
two parallel surfaces $S_{\varepsilon}^{\pm}$.\label{fig:geometry}}
\end{figure}

As a general example, let $\Omega_{1}\subset\mathbb{R}^{n}$ with
$n=1$, $2$ or $3$ be a finite spatial region enclosed by a boundary
$S=\partial\Omega_{1}$ with outward pointing normal $\mathbf{n}$
and let $\Omega_{2}=\mathbb{R}^{n}\setminus\overline{\Omega}_{1}$ be the
exterior region (see Figure \ref{fig:geometry}, left side).
Assuming a potential energy and an electron mass, respectively, of
the form: 
\begin{equation}
V\left(\mathbf{r}\right)=\begin{cases}
0 & \mathbf{r}\in\Omega_{1};\\
V & \mathbf{r}\in\Omega_{2},
\end{cases}\quad m\left(\mathbf{r}\right)=\begin{cases}
m_{1} & \mathbf{r}\in\Omega_{1};\\
m_{2} & \mathbf{r}\in\Omega_{2},
\end{cases}\label{eq:potentialAndMass}
\end{equation}
with $V$, $m_1$ and $m_2$ constants, equation (\ref{eq:scalarHelmholtz}) can be rewritten as:
\begin{equation}
\begin{cases}
\Delta\psi\left(\mathbf{r}\right)+k_{1}^{2}\psi\left(\mathbf{r}\right)=0 & \mathbf{r}\in\Omega_{1};\\
\Delta\psi\left(\mathbf{r}\right)+k_{2}^{2}\psi\left(\mathbf{r}\right)=0 & \mathbf{r}\in\Omega_{2},
\end{cases}\label{eq:system}
\end{equation}
where:
\begin{equation}
k_{1}^{2}=\frac{2m_{1}}{\hbar^{2}}E,\quad k_{2}^{2}=\frac{2m_{2}}{\hbar^{2}}\left(E-V\right).\label{eq:wavenumbers}
\end{equation}

Let us represent the total electron wave function $\psi\left(\mathbf{r}\right)$
as follows:
\begin{equation}
\psi\left(\mathbf{r}\right)\equiv\begin{cases}
\psi_{1}\left(\mathbf{r}\right) & \mathbf{r}\in\Omega_{1};\\
\psi_{2}\left(\mathbf{r}\right) & \mathbf{r}\in\Omega_{2}.
\end{cases}\label{eq:wavefunctionDefinition}
\end{equation}
With this notation, (\ref{eq:system}) becomes:
\begin{equation}
\Delta\psi_{j}\left(\mathbf{r}\right)+k_{j}^{2}\psi_{j}\left(\mathbf{r}\right)=0\quad j=1,2.\label{eq:compactEquation}
\end{equation}
Denoting by $g_{1}\left(\mathbf{r},\mathbf{r}'\right)$ and
$g_{2}\left(\mathbf{r},\mathbf{r}'\right)$ the free-space
Green functions in the two regions, we have:
\begin{equation}
\Delta g_{j}\left(\mathbf{r},\mathbf{r}'\right)+k_{j}^{2}g_{j}\left(\mathbf{r},\mathbf{r}'\right)=-\delta\left(\mathbf{r}-\mathbf{r}'\right)\quad j=1,2.\label{eq:greenEqu}
\end{equation}
If we make the replacement $\mathbf{r}\leftrightarrow\mathbf{r}'$
in (\ref{eq:compactEquation}) and (\ref{eq:greenEqu}), multiply
the first expression by $g_{j}\left(\mathbf{r},\mathbf{r}'\right)$,
the second by $\psi_{j}\left(\mathbf{r}'\right)$ and finally
compute the difference between the two, we arrive at:
\begin{align}
g_{j}\Delta'\psi_{j}\left(\mathbf{r}'\right)-\psi_{j}\left(\mathbf{r}'\right)\Delta'g_{j}=\psi_{j}\left(\mathbf{r}'\right)\delta\left(\mathbf{r}-\mathbf{r}'\right) & \quad j=1,2,\label{eq:difference}
\end{align}
where the argument $\left(\mathbf{r},\mathbf{r}'\right)$
of the Green function has been suppressed for brevity. Let us first
consider the case $\mathbf{r}\notin S$. By performing the integration
of (\ref{eq:difference}) in $d\mathbf{r}'$ over the volume $\Omega_{j}\setminus B_{\varrho}\left(\mathbf{r}\right)$,
where $B_{\varrho}\left(\mathbf{r}\right)$ is a ball of radius
$\varrho$ and center $\mathbf{r}\in\Omega_{j}$, making use of
the following Green's identity: 
\begin{equation}
\int_{\Omega}d\mathbf{r}'\left[g\Delta'\psi\left(\mathbf{r}'\right)-\psi\left(\mathbf{r}'\right)\Delta'g\right]=\int_{\partial\Omega}d\mathbf{r}'\left[g\frac{\partial\psi\left(\mathbf{r}'\right)}{\partial n'}-\psi\left(\mathbf{r}'\right)\frac{\partial g}{\partial n'}\right],\label{eq:GreenId}
\end{equation}
where $\partial/\partial n'=\mathbf{n}'\cdot\nabla'$ is the derivative
with respect to the outward pointing normal to the integration surface
computed at $\mathbf{r}'$, and taking the limit $\varrho\rightarrow0$,
we get:
\begin{equation}
\int_{S}d\mathbf{r}'\left[g_{j}\frac{\partial\psi_{j}\left(\mathbf{r}'\right)}{\partial n'}-\psi_{j}\left(\mathbf{r}'\right)\frac{\partial g_{j}}{\partial n'}\right]\mp\lim_{\varrho\rightarrow0}\int_{\partial B_{\varrho}\left(\mathbf{r}\right)}d\mathbf{r}'\left[g_{j}\frac{\partial\psi_{j}\left(\mathbf{r}'\right)}{\partial n'}-\psi_{j}\left(\mathbf{r}'\right)\frac{\partial g_{j}}{\partial n'}\right]=0\quad j=1,2.\label{eq:boundaryEquationDerivation}
\end{equation}
As it is shown in \ref{sec:A}, the second term in (\ref{eq:boundaryEquationDerivation})
reduces to: 
\begin{equation}
\lim_{\varrho\rightarrow0}\int_{\partial B_{\varrho}\left(\mathbf{r}\right)}d\mathbf{r}'\left[g_{j}\frac{\partial\psi_{j}\left(\mathbf{r}'\right)}{\partial n'}-\psi_{j}\left(\mathbf{r}'\right)\frac{\partial g_{j}}{\partial n'}\right]=\psi_{j}\left(\mathbf{r}\right)\quad j=1,2,\label{eq:sphericalIntegral}
\end{equation}
so that we are left with:
\begin{equation}
{\displaystyle \begin{cases}
\psi_{1}\left(\mathbf{r}\right)=\int_{S}d\mathbf{r}'\left[g_{1}\frac{\partial\psi_{1}\left(\mathbf{r}'\right)}{\partial n'}-\psi_{1}\left(\mathbf{r}'\right)\frac{\partial g_{1}}{\partial n'}\right] & \mathbf{r}\in\Omega_{1};\\
\psi_{2}\left(\mathbf{r}\right)=\int_{S}d\mathbf{r}'\left[\psi_{2}\left(\mathbf{r}'\right)\frac{\partial g_{2}}{\partial n'}-g_{2}\frac{\partial\psi_{2}\left(\mathbf{r}'\right)}{\partial n'}\right] & \mathbf{r}\in\Omega_{2}.
\end{cases}}\label{eq:SolSystem}
\end{equation}
In deriving the second equation in (\ref{eq:SolSystem}), the Sommerfeld radiation condition (\ref{eq:sommerfeld}) has been used
in order to neglect the contribution at infinity. Taking $f\left(\cdot\right)$
to represent $\psi\left(\cdot\right)$, $g\left(\mathbf{r},\cdot\right)$
and their normal derivative, the evaluation of the integrand functions
at any point $\mathbf{r}_{S}$ on the boundary should be conceived
as follows\footnote{In order to avoid abuse of notation, the normal derivatives must be
computed on the surfaces defined by $\mathbf{r}_{S}-\varepsilon\mathbf{n}$
and $\mathbf{r}_{S}+\varepsilon\mathbf{n}$, respectively.}:
\begin{equation}
f_{1}\left(\mathbf{r}_{S}\right)\equiv\lim_{\varepsilon\rightarrow0}f_{1}\left(\mathbf{r}_{S}-\varepsilon\mathbf{n}\right),\quad f_{2}\left(\mathbf{r}_{S}\right)\equiv\lim_{\varepsilon\rightarrow0}f_{2}\left(\mathbf{r}_{S}+\varepsilon\mathbf{n}\right).\label{eq:boundaryLimits}
\end{equation}
The same strategy can be used to obtain the limiting values of (\ref{eq:SolSystem})
for $\mathbf{r}\in S$. In one dimension, where the boundary integrals
are replaced by point evaluations, this poses no problem. Conversely,
owing to the singularity of the Green functions, special care must
be taken to address the two-dimensional and three-dimensional cases.
In particular, by deforming the boundary integrals so that $\mathbf{r}$
still resides inside $\Omega_{j}$, it can be shown that (see
\ref{sec:B})\footnote{An alternative derivation would require to apply again (\ref{eq:difference})
and (\ref{eq:GreenId}) integrating over $\Omega_{j}\setminus\left[\Omega_{j}\cap B_{\varrho}\left(\mathbf{r}_{S}\right)\right]$.}:
\begin{equation}
{\displaystyle \begin{cases}
\frac{\psi_{1}\left(\mathbf{r}\right)}{2}=\fint_{S}d\mathbf{r}'\left[g_{1}\frac{\partial\psi_{1}\left(\mathbf{r}'\right)}{\partial n'}-\psi_{1}\left(\mathbf{r}'\right)\frac{\partial g_{1}}{\partial n'}\right] & \mathbf{r}\in S;\\
\frac{\psi_{2}\left(\mathbf{r}\right)}{2}=\fint_{S}d\mathbf{r}'\left[\psi_{2}\left(\mathbf{r}'\right)\frac{\partial g_{2}}{\partial n'}-g_{2}\frac{\partial\psi_{2}\left(\mathbf{r}'\right)}{\partial n'}\right] & \mathbf{r}\in S,
\end{cases}}\label{eq:SolSystemRegularized}
\end{equation}
where the symbol $\fint$ stands for the Cauchy principal value integral.

We now introduce the boundary conditions for the wave function. First,
let us define: 
\begin{equation}
\chi_{j}\left(\mathbf{r}_{S}\right)\equiv\frac{1}{m_{j}}\frac{\partial\psi_{j}\left(\mathbf{r}_{S}\right)}{\partial n}\quad j=1,2.\label{eq:normalDerivatives}
\end{equation}
With this convention, and taking into account (\ref{eq:boundaryLimits}),
the continuity of the electron wave function and of its normal derivative
at the boundary $S$ can be expressed as:
\begin{equation}
{\displaystyle \begin{cases}
\delta\psi\left(\mathbf{r}_{S}\right)\equiv\psi_{2}\left(\mathbf{r}_{S}\right)-\psi_{1}\left(\mathbf{r}_{S}\right)=0\quad\Rightarrow & \psi_{2}\left(\mathbf{r}_{S}\right)=\psi_{1}\left(\mathbf{r}_{S}\right)\equiv\psi\left(\mathbf{r}_{S}\right);\\
\delta\chi\left(\mathbf{r}_{S}\right)\equiv\chi_{2}\left(\mathbf{r}_{S}\right)-\chi_{1}\left(\mathbf{r}_{S}\right)=0\quad\Rightarrow & \chi_{2}\left(\mathbf{r}_{S}\right)=\chi_{1}\left(\mathbf{r}_{S}\right)\equiv\chi\left(\mathbf{r}_{S}\right).
\end{cases}}\label{eq:boundaryCondSystem}
\end{equation}
System (\ref{eq:SolSystemRegularized}) is then reduced to:
\begin{equation}
{\displaystyle \begin{cases}
\frac{\psi\left(\mathbf{r}_{S}\right)}{2}+\fint_{S}d\mathbf{r}_{S}'\left[\frac{\partial g_{1}\left(\mathbf{r}_{S},\mathbf{r}_{S}'\right)}{\partial n'}\psi\left(\mathbf{r}_{S}'\right)-m_{1}g_{1}\left(\mathbf{r}_{S},\mathbf{r}_{S}'\right)\chi\left(\mathbf{r}_{S}'\right)\right]=0;\\
\frac{\psi\left(\mathbf{r}_{S}\right)}{2}+\fint_{S}d\mathbf{r}_{S}'\left[-\frac{\partial g_{2}\left(\mathbf{r}_{S},\mathbf{r}_{S}'\right)}{\partial n'}\psi\left(\mathbf{r}_{S}'\right)+m_{2}g_{2}\left(\mathbf{r}_{S},\mathbf{r}_{S}'\right)\chi\left(\mathbf{r}_{S}'\right)\right]=0
\end{cases}}\label{eq:SolSystemRegularizedSimplified}
\end{equation}
and, by subtracting the two equations, we arrive at:
\begin{equation}
\fint_{S}d\mathbf{r}_{S}'\left[\frac{\partial g_{1}\left(\mathbf{r}_{S},\mathbf{r}_{S}'\right)}{\partial n'}+\frac{\partial g_{2}\left(\mathbf{r}_{S},\mathbf{r}_{S}'\right)}{\partial n'}\right]\psi\left(\mathbf{r}_{S}'\right)-\fint_{S}d\mathbf{r}_{S}'\left[m_{1}g_{1}\left(\mathbf{r}_{S},\mathbf{r}_{S}'\right)+m_{2}g_{2}\left(\mathbf{r}_{S},\mathbf{r}_{S}'\right)\right]\chi\left(\mathbf{r}_{S}'\right)=0.\label{eq:firstEqu}
\end{equation}

Let us now consider the following families of parallel surfaces: 
\begin{equation}
S_{\varepsilon}^{-}\equiv\left\{ \mathbf{r}\in\Omega_{1}:\:\mathbf{r}=\mathbf{r}_{S}-\varepsilon\mathbf{n},\:\mathbf{r}_{S}\in S\right\} ,\quad S_{\varepsilon}^{+}\equiv\left\{ \mathbf{r}\in\Omega_{2}:\:\mathbf{r}=\mathbf{r}_{S}+\varepsilon\mathbf{n},\:\mathbf{r}_{S}\in S\right\} ,\label{eq:parallelSurfacesFamily}
\end{equation}
denote by $\mathbf{n}_{\mp}$ the corresponding outward pointing
normals, by $\Omega_{\varepsilon}^{-}\subset\Omega_{1}$ the inner
volume with respect to $S_{\varepsilon}^{-}$ and by $\Omega_{\varepsilon}^{+}\subset\Omega_{2}$
the outer volume with respect to $S_{\varepsilon}^{+}$ (see Figure
\ref{fig:geometry}, right side). On computing the normal derivative
of (\ref{eq:SolSystem}) at $S_{\varepsilon}^{\mp}$, respectively,
we get:
\begin{equation}
{\displaystyle \begin{cases}
\frac{\partial\psi_{1}\left(\mathbf{r}\right)}{\partial n_{-}}=\int_{S}d\mathbf{r}'\left[\frac{\partial g_{1}}{\partial n_{-}}\frac{\partial\psi_{1}\left(\mathbf{r}'\right)}{\partial n'}-\psi_{1}\left(\mathbf{r}'\right)\frac{\partial^{2}g_{1}}{\partial n_{-}\partial n'}\right] & \mathbf{r}\in S_{\varepsilon}^{-};\\
\frac{\partial\psi_{2}\left(\mathbf{r}\right)}{\partial n_{+}}=\int_{S}d\mathbf{r}'\left[\psi_{2}\left(\mathbf{r}'\right)\frac{\partial^{2}g_{2}}{\partial n_{+}\partial n'}-\frac{\partial g_{2}}{\partial n_{+}}\frac{\partial\psi_{2}\left(\mathbf{r}'\right)}{\partial n'}\right] & \mathbf{r}\in S_{\varepsilon}^{+}.
\end{cases}}\label{eq:systemNormDer}
\end{equation}
Dividing the first equation by $m_{1}$, the second by $m_{2}$ and
taking the difference between the two under the limit $\varepsilon\rightarrow0$
gives (see \ref{sec:C}):
\begin{equation}
\int_{S}d\mathbf{r}_{S}'\left[\frac{1}{m_{1}}\frac{\partial^{2}g_{1}\left(\mathbf{r}_{S},\mathbf{r}_{S}'\right)}{\partial n\partial n'}+\frac{1}{m_{2}}\frac{\partial^{2}g_{2}\left(\mathbf{r}_{S},\mathbf{r}_{S}'\right)}{\partial n\partial n'}\right]\psi\left(\mathbf{r}_{S}'\right)-\fint_{S}d\mathbf{r}_{S}'\left[\frac{\partial g_{1}\left(\mathbf{r}_{S},\mathbf{r}_{S}'\right)}{\partial n}+\frac{\partial g_{2}\left(\mathbf{r}_{S},\mathbf{r}_{S}'\right)}{\partial n}\right]\chi\left(\mathbf{r}_{S}'\right)=0.\label{eq:secondEqu}
\end{equation}
It is fundamental to keep in mind that the first integral in (\ref{eq:secondEqu})
is hypersingular and does not exist as Cauchy principal value. 

Equations (\ref{eq:firstEqu}) and (\ref{eq:secondEqu}) can be rewritten
more concisely as:
\begin{equation}
\hat{\mathbf{H}}\left[\begin{array}{c}
\psi\\
\chi
\end{array}\right]\left(\mathbf{r}_{S}\right)=0,\label{eq:matrixEqu}
\end{equation}
where the operator $\hat{\mathbf{H}}$ is expressed in matrix
form (here we adapt the notation of \cite{Nedelec2001} to the present scenario): 
\begin{equation}
\hat{\mathbf{H}}\equiv\left(\begin{array}{cc}
-\hat{D} & \hat{S}\\
-\hat{N} & \hat{D}^{\dagger}
\end{array}\right)\label{eq:operatorH}
\end{equation}
with entries defined as boundary integral operators over an arbitrary
wave function $f\left(\mathbf{r}\right)$:
\begin{align}
\hat{S}\left[f\right]\left(\mathbf{r}_{S}\right) & \equiv\fint_{S}d\mathbf{r}_{S}'\left[m_{1}g_{1}\left(\mathbf{r}_{S},\mathbf{r}_{S}'\right)+m_{2}g_{2}\left(\mathbf{r}_{S},\mathbf{r}_{S}'\right)\right]f\left(\mathbf{r}_{S}'\right);\\
\hat{D}\left[f\right]\left(\mathbf{r}_{S}\right) & \equiv\fint_{S}d\mathbf{r}_{S}'\left[\frac{\partial g_{1}\left(\mathbf{r}_{S},\mathbf{r}_{S}'\right)}{\partial n'}+\frac{\partial g_{2}\left(\mathbf{r}_{S},\mathbf{r}_{S}'\right)}{\partial n'}\right]f\left(\mathbf{r}_{S}'\right);\\
\hat{D}^{\dagger}\left[f\right]\left(\mathbf{r}_{S}\right) & \equiv\fint_{S}d\mathbf{r}_{S}'\left[\frac{\partial g_{1}\left(\mathbf{r}_{S},\mathbf{r}_{S}'\right)}{\partial n}+\frac{\partial g_{2}\left(\mathbf{r}_{S},\mathbf{r}_{S}'\right)}{\partial n}\right]f\left(\mathbf{r}_{S}'\right);\\
\hat{N}\left[f\right]\left(\mathbf{r}_{S}\right) & \equiv\int_{S}d\mathbf{r}_{S}'\left[\frac{1}{m_{1}}\frac{\partial^{2}g_{1}\left(\mathbf{r}_{S},\mathbf{r}_{S}'\right)}{\partial n\partial n'}+\frac{1}{m_{2}}\frac{\partial^{2}g_{2}\left(\mathbf{r}_{S},\mathbf{r}_{S}'\right)}{\partial n\partial n'}\right]f\left(\mathbf{r}_{S}'\right).
\end{align}

To summarize, the solution of the original Schr\"{o}dinger equation in
the two regions has been rewritten through (\ref{eq:SolSystem}) and
(\ref{eq:boundaryCondSystem}) as an integral expression involving
the values of the functions $\psi$ and $\chi$ at the boundary:
\begin{equation}
{\displaystyle \begin{cases}
\psi_{1}\left(\mathbf{r}\right)=\int_{S}d\mathbf{r}_{S}'\left[m_{1}g_{1}\left(\mathbf{r},\mathbf{r}_{S}'\right)\chi\left(\mathbf{r}_{S}'\right)-\frac{\partial g_{1}\left(\mathbf{r},\mathbf{r}_{S}'\right)}{\partial n'}\psi\left(\mathbf{r}_{S}'\right)\right] & \mathbf{r}\in\Omega_{1};\\
\psi_{2}\left(\mathbf{r}\right)=\int_{S}d\mathbf{r}_{S}'\left[\frac{\partial g_{2}\left(\mathbf{r},\mathbf{r}_{S}'\right)}{\partial n'}\psi\left(\mathbf{r}_{S}'\right)-m_{2}g_{2}\left(\mathbf{r},\mathbf{r}_{S}'\right)\chi\left(\mathbf{r}_{S}'\right)\right] & \mathbf{r}\in\Omega_{2}.
\end{cases}}\label{eq:BEMsolution}
\end{equation}
In (\ref{eq:matrixEqu}), the boundary restrictions $\psi\left(\mathbf{r}_{S}\right)$
and $\chi\left(\mathbf{r}_{S}\right)$ are found to span the null space
of the matrix integral operator (\ref{eq:operatorH}). It is important to note that the $\hat{S}$, $\hat{D}$, $\hat{D}^{\dagger}$ and $\hat{N}$ operators are coercive \cite{Steinbach2008} but they may lack injectivity for some discrete values of the electron energy depending on the geometry and physical parameters of the system: those energies constitute the bound portion of the spectrum. The bound states of the quantum problem are intimately related to the resonant modes of the corresponding Helmholtz problem, with the presence in both cases of a non-trivial null space of the BEM operator. 

\subsection{Discretization of the operators\label{subsec:Discretization}}

In order to solve numerically the above derived integral equations,
the boundary $S$ is discretized into a collection of simplices $\left\{ S_{n}\right\} $
(segments and triangles in two and three dimensions, respectively).
We then expand the unknowns $\psi$ and $\chi$ in (\ref{eq:matrixEqu})
on a set of node-based basis functions $\left\{ f_{j}\right\} $ as
follows:
\begin{equation}
\psi\left(\mathbf{r}_{S}'\right)=\sum_{j}\alpha_{j}\,f_{j}\left(\mathbf{r}_{S}'\right);\quad\chi\left(\mathbf{r}_{S}'\right)=\mu^{-1}\sum_{j}\beta_{j}\,f_{j}\left(\mathbf{r}_{S}'\right),\label{eq:bfExpansion}
\end{equation}
being $\mu$ a dimensionless constant with the same order of magnitude
as the electron mass, used to avoid scaling issues in the numeric
computation. The $j$-th basis function is defined on the set of simplices
$\left\{ S_{n}\right\} $ that share the $j$-th mesh node, hereinafter
referred to as $\left\{ n\in j\right\} $, and vanishes out of its
defining domain, so that:
\begin{equation}
\int_{S}d\mathbf{r}_{S}'f_{j}\left(\mathbf{r}_{S}'\right)=\sum_{n\in j}\int_{S_{n}}d\mathbf{r}'f_{j}^{n}\left(\mathbf{r}'\right),\label{eq:bfRestriction}
\end{equation}
with $f_{j}^{n}$ representing the restriction of the basis function
to the $n$-th simplex. Following the Galerkin approach \cite{Sutradhar2008},
we multiply equation (\ref{eq:matrixEqu}) by $f_{i}\left(\mathbf{r}_{S}\right)$
and integrate over $S$ to obtain:
\begin{equation}
\sum_{j}\mathbf{H}_{ij}\left[\begin{array}{c}
\alpha_{j}\\
\beta_{j}
\end{array}\right]=0,\label{eq:matrixEquDiscrete}
\end{equation}
where:
\begin{equation}
\mathbf{H}_{ij}\equiv\left(\begin{array}{cc}
-D_{ij} & \mu^{-1}S_{ij}\\
-\mu \, N_{ij} & D_{ij}^{\dagger}
\end{array}\right)\label{eq:operatorHdiscrete}
\end{equation}
and the discrete boundary operators are given by:
\begin{align}
S_{ij} & \equiv\sum_{m\in i}\,\sum_{n\in j}\int_{S_{m}}d\mathbf{r}\fint_{S_{n}}d\mathbf{r}'\left[m_{1}g_{1}\left(\mathbf{r},\mathbf{r}'\right)+m_{2}g_{2}\left(\mathbf{r},\mathbf{r}'\right)\right]f_{i}^{m}\left(\mathbf{r}\right)f_{j}^{n}\left(\mathbf{r}'\right);\label{eq:singleLayer}\\
D_{ij} & \equiv\sum_{m\in i}\,\sum_{n\in j}\int_{S_{m}}d\mathbf{r}\fint_{S_{n}}d\mathbf{r}'\left[\frac{\partial g_{1}\left(\mathbf{r},\mathbf{r}'\right)}{\partial n'}+\frac{\partial g_{2}\left(\mathbf{r},\mathbf{r}'\right)}{\partial n'}\right]f_{i}^{m}\left(\mathbf{r}\right)f_{j}^{n}\left(\mathbf{r}'\right);\\
D_{ij}^{\dagger} & \equiv\sum_{m\in i}\,\sum_{n\in j}\int_{S_{m}}d\mathbf{r}\fint_{S_{n}}d\mathbf{r}'\left[\frac{\partial g_{1}\left(\mathbf{r},\mathbf{r}'\right)}{\partial n}+\frac{\partial g_{2}\left(\mathbf{r},\mathbf{r}'\right)}{\partial n}\right]f_{i}^{m}\left(\mathbf{r}\right)f_{j}^{n}\left(\mathbf{r}'\right);\\
N_{ij} & \equiv\sum_{m\in i}\,\sum_{n\in j}\int_{S_{m}}d\mathbf{r}\int_{S_{n}}d\mathbf{r}'\left[\frac{1}{m_{1}}\frac{\partial^{2}g_{1}\left(\mathbf{r},\mathbf{r}'\right)}{\partial n\partial n'}+\frac{1}{m_{2}}\frac{\partial^{2}g_{2}\left(\mathbf{r},\mathbf{r}'\right)}{\partial n\partial n'}\right]f_{i}^{m}\left(\mathbf{r}\right)f_{j}^{n}\left(\mathbf{r}'\right).\label{eq:hypersingular}
\end{align}
When the simplices $S_{m}$ and $S_{n}$ do not share any vertex,
the matrix entries (\ref{eq:singleLayer})-(\ref{eq:hypersingular})
can be easily computed by Gauss-Legendre quadrature rules \cite{Abramowitz1972}.
Conversely, owing to the singularity of the Green functions and their
normal derivatives, most integrations over coincident and adjacent
elements require the use of regularization techniques (see, for instance,
\cite{Wilton1984}). Following the variational formulation proposed
in \cite{Nedelec2001} and \cite{Steinbach2008}, the hypersingular
matrix (\ref{eq:hypersingular}) may be replaced by a discrete version
of the bilinear form induced by the corresponding single layer potential,
which proves similar to (\ref{eq:singleLayer}) and easier to deal
with. In \ref{sec:D}, quasi-closed-form expressions are
provided for the coincident integrations appearing throughout (\ref{eq:singleLayer})-(\ref{eq:hypersingular})
in the two-dimensional case with first-order basis functions. To the
best of our knowledge, these formulas are applied here for the first
time. 

Once the above matrices are computed, the sets of expansion coefficients
$\left\{ \alpha_{j}\right\} $ and $\left\{ \beta_{j}\right\} $ can
be estimated by solving (\ref{eq:matrixEquDiscrete}) numerically,
and this in turn leads to the determination of the boundary unknowns
$\psi\left(\mathbf{r}_{S}\right)$ and $\chi\left(\mathbf{r}_{S}\right)$
via (\ref{eq:bfExpansion}). Since the matrix entries depend parametrically
on the energy $E$ of the electron, root-finding methods must be employed
to localize the bound states, seeking for those specific eigenenergies
that lead to a vanishing determinant of the block matrix (\ref{eq:operatorHdiscrete}).
Finally, from (\ref{eq:BEMsolution}), (\ref{eq:bfExpansion}) and
(\ref{eq:bfRestriction}), we arrive at the BEM solution:
\begin{equation}
{\displaystyle \begin{cases}
\psi_{1}\left(\mathbf{r}\right)=\sum_{j}\sum_{n\in j}\int_{S_{n}}d\mathbf{r}'\left[\beta_{j}\frac{m_{1}}{\mu}g_{1}\left(\mathbf{r},\mathbf{r}'\right)-\alpha_{j}\frac{\partial g_{1}\left(\mathbf{r},\mathbf{r}'\right)}{\partial n'}\right]\,f_{j}^{n}\left(\mathbf{r}'\right) & \mathbf{r}\in\Omega_{1};\\
\psi_{2}\left(\mathbf{r}\right)=\sum_{j}\sum_{n\in j}\int_{S_{n}}d\mathbf{r}'\left[\alpha_{j}\frac{\partial g_{2}\left(\mathbf{r},\mathbf{r}'\right)}{\partial n'}-\beta_{j}\frac{m_{2}}{\mu}g_{2}\left(\mathbf{r},\mathbf{r}'\right)\right]\,f_{j}^{n}\left(\mathbf{r}'\right) & \mathbf{r}\in\Omega_{2},
\end{cases}}
\end{equation}
which can be more usefully expressed as:
\begin{equation}
{\displaystyle \begin{cases}
\psi_{1}\left(\mathbf{r}\right)=\sum_{n}\int_{S_{n}}d\mathbf{r}'\sum_{j\in n}\left[\beta_{j}\frac{m_{1}}{\mu}g_{1}\left(\mathbf{r},\mathbf{r}'\right)-\alpha_{j}\frac{\partial g_{1}\left(\mathbf{r},\mathbf{r}'\right)}{\partial n'}\right]\,f_{j}^{n}\left(\mathbf{r}'\right) & \mathbf{r}\in\Omega_{1};\\
\psi_{2}\left(\mathbf{r}\right)=\sum_{n}\int_{S_{n}}d\mathbf{r}'\sum_{j\in n}\left[\alpha_{j}\frac{\partial g_{2}\left(\mathbf{r},\mathbf{r}'\right)}{\partial n'}-\beta_{j}\frac{m_{2}}{\mu}g_{2}\left(\mathbf{r},\mathbf{r}'\right)\right]\,f_{j}^{n}\left(\mathbf{r}'\right) & \mathbf{r}\in\Omega_{2},
\end{cases}}\label{eq:BEMwavefunction}
\end{equation}
being $\left\{ j\in n\right\} $ the set of mesh nodes that belong
to the $n$-th simplex.

\subsection{Examples and comparisons\label{subsec:boundExamples}}

\begin{figure}
\begin{centering}
\includegraphics[width=1\textwidth]{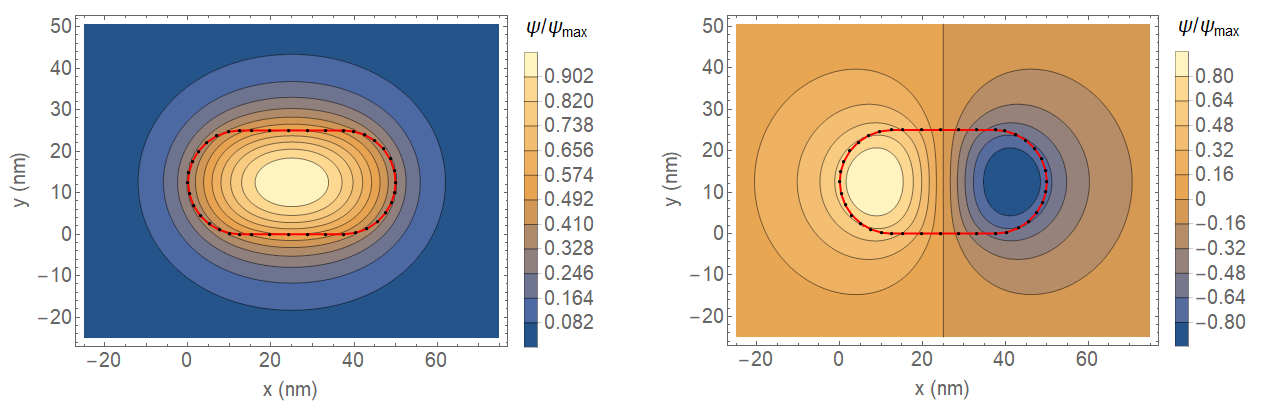}
\par\end{centering}
\caption{Contour plots of the two bound electron wave functions in a stadium-shaped
structure of size $50\times25\:\mathrm{nm}^{2}$ with $m_{1}=m_{2}=0.0665\,m_{e}$
and $V=10\:\mathrm{meV}$. The wave functions are computed by the proposed BEM
formulation via (\ref{eq:matrixEquDiscrete}) and (\ref{eq:BEMwavefunction})
using a mesh of $40$ elements, first-order basis functions and a
$10$ points Gauss-Legendre quadrature for the numerical integrations.\label{fig:stadiumStates}}
\end{figure}

\begin{figure}
\begin{centering}
\includegraphics[width=0.6\textwidth]{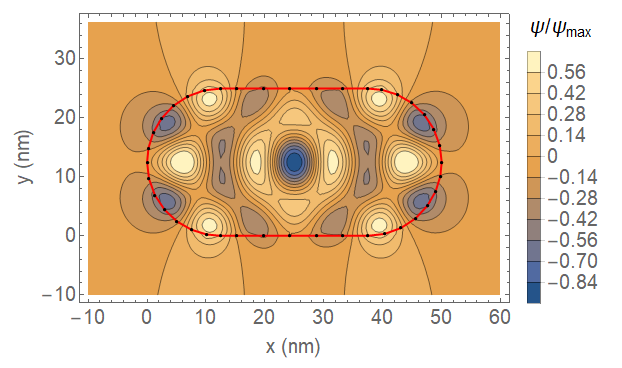}
\par\end{centering}
\caption{Contour plot of the electron wave function in a stadium-shaped structure
of size $50\times25\:\mathrm{nm}^{2}$ with $m_{1}=m_{2}=0.0665\,m_{e}$,
$E=184.4\:\mathrm{meV}$ and $V=190\:\mathrm{meV}$. The wave function is computed by
the proposed BEM formulation via (\ref{eq:matrixEquDiscrete}) and
(\ref{eq:BEMwavefunction}) using a mesh of $40$ elements, first-order
basis functions and a $10$ points Gauss-Legendre quadrature for the
numerical integrations.\label{fig:stadiumExcitedState}}
\end{figure}

With reference to \cite{Knipp1996}, let us first consider a stadium-shaped boundary of size $50\times25\:\mathrm{nm}^{2}$ with a potential offset $V=10$ meV and take the electron effective mass to be $0.0665\,m_{e}$ in both the inner and outer regions, where $m_{e}\approx9.11\times10^{-31}$ Kg represents the electron rest mass. The contour plots of the two bound electron states computed by the proposed BEM are shown in Figure \ref{fig:stadiumStates}, whereas the corresponding
energies are reported in Table \ref{tab:energies} for different numbers
of mesh elements $\mathcal{N}$. From this analysis, the relative error of the calculated energies is found to decrease as $O(\mathcal{N}^{-2})$. As a further comparison, the contour plot of the excited electron
state at $184.4$ meV, computed using $V=190$ meV, is displayed in
Figure \ref{fig:stadiumExcitedState}. The results are in good agreement with those presented in \cite{Knipp1996}.
\begin{table}
\begin{centering}
\begin{tabular}{|c|c|c|}
\hline 
number of mesh elements & first energy level (meV) & second energy level (meV)\tabularnewline
\hhline{|=|=|=|}
16 & 4.8494 & 8.7556\tabularnewline
\hline 
24 & 4.8255 & 8.6924\tabularnewline
\hline 
32 & 4.8128 & 8.6588\tabularnewline
\hline 
40 & 4.8090 & 8.6496\tabularnewline
\hline 
50 & 4.8074 & 8.6443\tabularnewline
\hline 
100 & 4.8023 & 8.6325\tabularnewline
\hline 
200 & 4.8021 & 8.6305 \tabularnewline
\hline 
\end{tabular}
\par\end{centering}
\caption{BEM-computed energies of the two bound electron states in a stadium-shaped
structure with the same parameters as in Figure \ref{fig:stadiumStates}.
The energies are obtained minimizing the function $\left|\det\mathbf{H}\left(E\right)\right|$
by standard root-finding algorithms. \label{tab:energies}}
\end{table}

\begin{figure}
\begin{centering}
\includegraphics[width=1\textwidth]{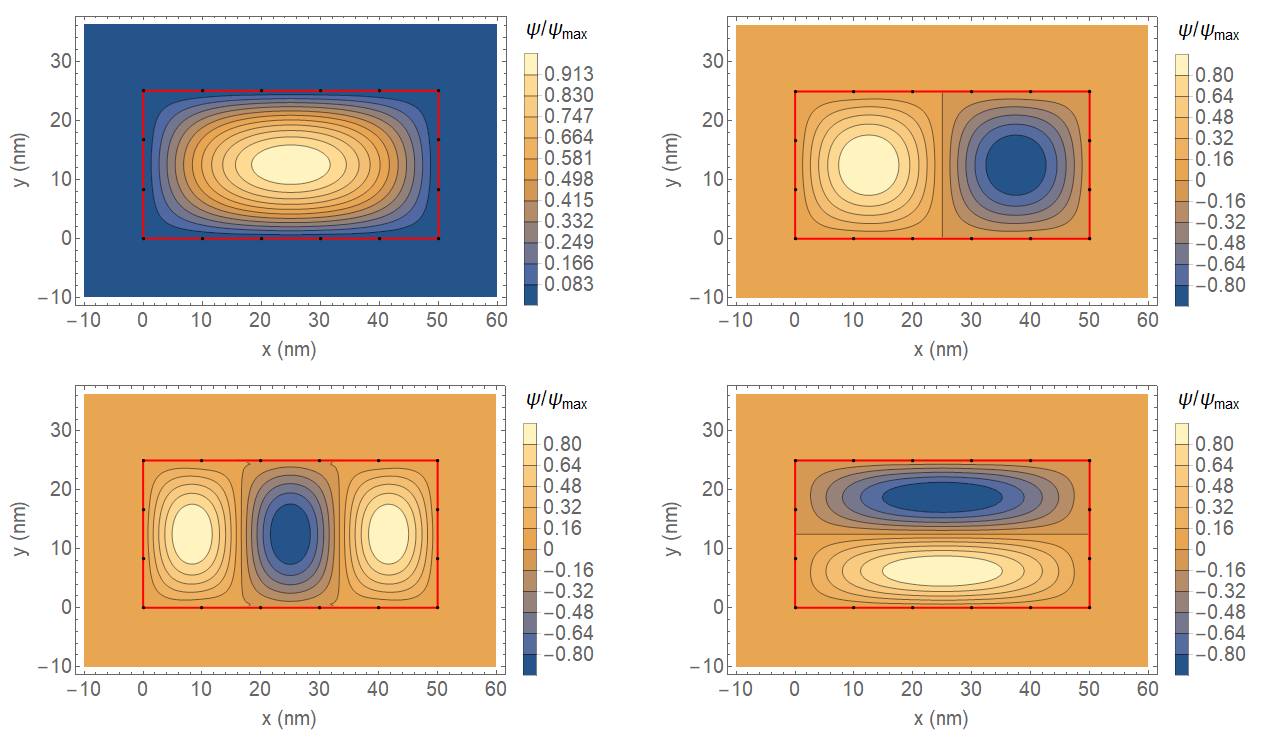}
\par\end{centering}
\caption{Contour plots of the first four bound electron wave functions in a
rectangular structure of size $50\times25\:\mathrm{nm}^{2}$ with
$m_{1}=m_{2}=0.0665\,m_{e}$ and $V\gg E$. The wave functions are
computed by the proposed BEM formulation via (\ref{eq:discretizedInfinitePotentialEquation})
and (\ref{eq:discretizedInfinitePotentialSolution}) using a mesh of $16$ elements, first-order
basis functions and a $10$ points Gauss-Legendre quadrature for the
numerical integrations.\label{fig:rectangleStates}}
\end{figure}

\begin{figure}
\begin{centering}
\includegraphics[width=0.68\textwidth]{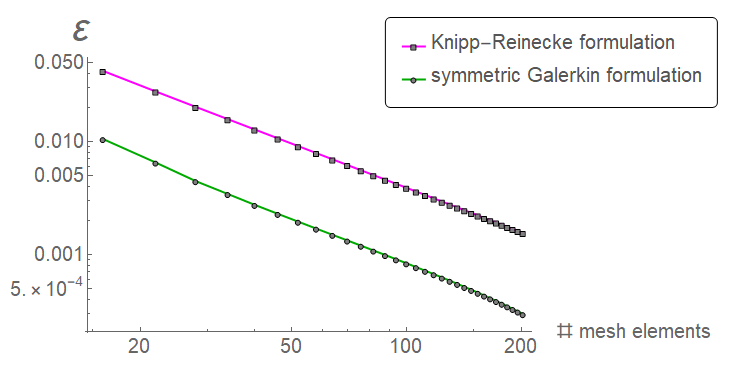}
\par\end{centering}
\caption{BEM error in the approximation of the first bound electron state inside a rectangular structure
as a function of the number of mesh elements for the same choice of
parameters adopted in Figure \ref{fig:rectangleStates}. The proposed symmetric Galerkin formulation is compared with that in \cite{Knipp1996}. \label{fig:ErrorVsMesh}}
\end{figure}

\begin{figure}
\begin{centering}
\includegraphics[width=0.86\textwidth]{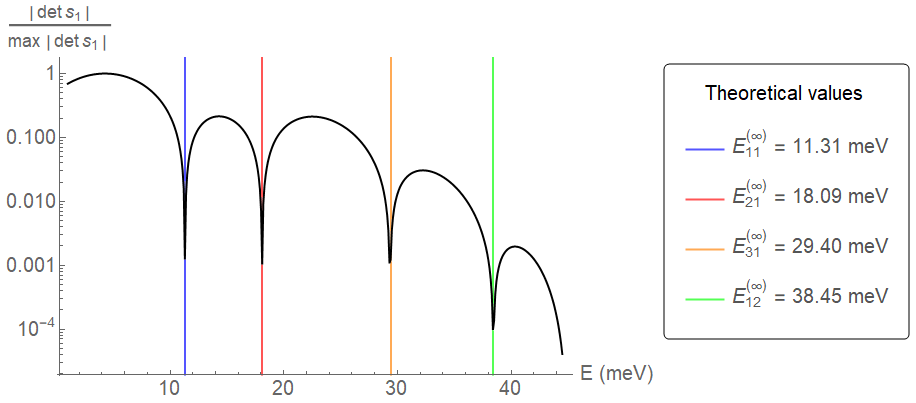}
\par\end{centering}
\caption{Determinant of the matrix (\ref{eq:discretizedInfinitePotentialOperator}) as a
function of the electron energy for the same choice of parameters
adopted in Figure \ref{fig:rectangleStates}.\label{fig:determinant}}
\end{figure}

The stadium is now replaced by a rectangular boundary of the same size. When the potential offset $V$
in (\ref{eq:potentialAndMass}) tends to infinity, the electron wave
function in the outer region vanishes and the Schr\"{o}dinger equation
for the inner region can be solved very easily by separation of variables.
Imposing the continuity of the wave function at the boundary and the
normalization condition, we obtain: 
\begin{equation}
{\displaystyle \begin{cases}
\psi_{1}^{(\infty)}\left(x,y\right)=\sqrt{\frac{2}{L_{x}}}\sin\left(\frac{n_{x}\pi x}{L_{x}}\right)\sqrt{\frac{2}{L_{y}}}\sin\left(\frac{n_{y}\pi y}{L_{y}}\right) & \left(x,y\right)\in\Omega_{1};\\
\psi_{2}^{(\infty)}\left(x,y\right)=0 & \left(x,y\right)\in\Omega_{2},
\end{cases}}\label{eq:analytic}
\end{equation}
where $n_{x}$, $n_{y}\in\mathbb{Z}$ are the quantum numbers and $L_{x}$, $L_{y}$ represent the sides of the rectangle. The
energy of the confined states can then be expressed analytically as
follows:
\begin{equation}
E_{n_{x}n_{y}}^{(\infty)}=\frac{\hbar^{2}\pi^{2}}{2m}\left(\frac{n_{x}^{2}}{L_{x}^{2}}+\frac{n_{y}^{2}}{L_{y}^{2}}\right).\label{eq:energies}
\end{equation}
In the present scenario, the infinite potential offset breaks the continuity of the normal derivative of the wave function across the boundary and leads to a vanishing Green function in the outer region, so that system (\ref{eq:SolSystem}) reduces to: 
\begin{equation}
{\displaystyle \begin{cases}
\psi_{1}\left(\mathbf{r}\right)=\int_{S}d\mathbf{r}_{S}'\,g_{1}\left(\mathbf{r},\mathbf{r}_{S}'\right)\frac{\partial\psi_{1}\left(\mathbf{r}_{S}'\right)}{\partial n'} & \mathbf{r}\in\Omega_{1};\\
\psi_{2}\left(\mathbf{r}\right)=0 & \mathbf{r}\in\Omega_{2}
\end{cases}}\label{eq:infinitePotentialEquation}
\end{equation}
and the boundary conditions (\ref{eq:boundaryCondSystem}) are replaced by:
\begin{equation}
\psi_{2}\left(\mathbf{r}_{S}\right)=\psi_{1}\left(\mathbf{r}_{S}\right)\equiv\psi\left(\mathbf{r}_{S}\right)=0.\label{eq:easyBoundayConditions}
\end{equation}
Despite the BEM equations derived in the previous sections no longer hold, we can still use (\ref{eq:infinitePotentialEquation}) and (\ref{eq:easyBoundayConditions}) to express a simplified boundary integral equation only involving the normal derivative of the wave function:
\begin{equation}
\hat{\mathsf{s}}_1 \left[\frac{\partial\psi_1}{\partial n}\right] \left(\mathbf{r}_{S}\right)\equiv\fint_{S}d\mathbf{r}_{S}'\,g_1\left(\mathbf{r}_{S},\mathbf{r}_{S}'\right)\frac{\partial\psi_1\left(\mathbf{r}_{S}'\right)}{\partial n'}=0,
\end{equation}
which is discretized as usual: 
\begin{equation}
\frac{\partial\psi_{1}\left(\mathbf{r}_{S}'\right)}{\partial n'}=\sum_{j}\beta_{j}\,f_{j}\left(\mathbf{r}_{S}'\right);\quad \sum_{j}(\mathsf{s}_1)_{ij} \, \beta_j =0;\label{eq:discretizedInfinitePotentialEquation}
\end{equation}
\begin{equation}
(\mathsf{s}_1)_{ij} \equiv \sum_{m\in i}\,\sum_{n\in j}\int_{S_{m}}d\mathbf{r}\fint_{S_{n}}d\mathbf{r}'g_1\left(\mathbf{r},\mathbf{r}'\right)f_{i}^{m}\left(\mathbf{r}\right)f_{j}^{n}\left(\mathbf{r}'\right);\label{eq:discretizedInfinitePotentialOperator}
\end{equation}
\begin{equation}
\psi_{1}\left(\mathbf{r}\right)=\sum_{n}\int_{S_{n}}d\mathbf{r}'\,g_{1}\left(\mathbf{r},\mathbf{r}'\right)\,\sum_{j\in n}\beta_{j}\,f_{j}^{n}\left(\mathbf{r}'\right) \quad \mathbf{r}\in\Omega_{1}.\label{eq:discretizedInfinitePotentialSolution}
\end{equation}
To check the BEM formulation against the above analytical example,
the electron wave function is computed by setting the energy $E$
in (\ref{eq:wavenumbers}) to be one of the values (\ref{eq:energies}). The contour plots of the wave
functions relative to the first four energy levels, i.e., $E_{11}^{(\infty)}$,
$E_{21}^{(\infty)}$, $E_{31}^{(\infty)}$ and $E_{12}^{(\infty)}$,
are displayed in Figure \ref{fig:rectangleStates} and can be shown
to match those of the analytic solutions (\ref{eq:analytic}),
not reported here for brevity. Table \ref{tab:error} details the error $\mathcal{E}$ in the reconstructed wave functions, expressed by the following integral:
\begin{equation}
\mathcal{E} \equiv \sqrt{\int_{\Omega_1}d\mathbf{r}\,\left| \psi_{1}^{(\infty)}\left(\mathbf{r}\right)-\psi_{1}\left(\mathbf{r}\right)\right|^2} \label{eq:errorBEM}
\end{equation}
In Figure \ref{fig:ErrorVsMesh}, formula (\ref{eq:errorBEM}) is evaluated for an increasing number of mesh elements to check the convergence of the BEM algorithm. As illustrated in Figure \ref{fig:determinant}, a further validation to the model is provided by comparing (\ref{eq:energies}) with the
energy values that lead to a local minimum of the function $\left|\det\mathsf{s}_1\left(E\right)\right|$,
where $\mathsf{s}_1$ is the matrix defined in (\ref{eq:discretizedInfinitePotentialOperator}).
\begin{table}
\begin{centering}
\begin{tabular}{|c||c|c|c|c|}
\hline 
\quad $n_x$ \quad & 1 & 2 & 3 & 1\tabularnewline
\hline 
\quad $n_y$ \quad & 1 & 1 & 1 & 2 \tabularnewline
\hline 
$\mathcal{E}$ & \qquad 0.01 \qquad & \qquad 0.02 \qquad & \qquad 0.04 \qquad & \qquad 0.03 \qquad\tabularnewline
\hline 
\end{tabular}
\par\end{centering}
\caption{Numeric error (\ref{eq:errorBEM}) of the BEM in the approximation of the wave functions (\ref{eq:analytic}) inside a rectangular structure with the same parameters as in Figure \ref{fig:rectangleStates}. \label{tab:error}}
\end{table}

\section{Scattering states\label{sec:Scattering}}

\subsection{Integral equations}

Considering again the arbitrary two-region system introduced in Section
\ref{sec:BoundStates}, let us assume:
\begin{equation}
V\left(\mathbf{r}\right)=\begin{cases}
V & \mathbf{r}\in\Omega_{1};\\
0 & \mathbf{r}\in\Omega_{2},
\end{cases}\quad m\left(\mathbf{r}\right)=\begin{cases}
m_{1} & \mathbf{r}\in\Omega_{1};\\
m_{2} & \mathbf{r}\in\Omega_{2},
\end{cases}\label{eq:scattPotentialAndMass}
\end{equation}
so that the wavenumbers in (\ref{eq:system}) are given by:
\begin{equation}
k_{1}^{2}=\frac{2m_{1}}{\hbar^{2}}\left(E-V\right),\quad k_{2}^{2}=\frac{2m_{2}}{\hbar^{2}}E.
\end{equation}
We then express $\psi\left(\mathbf{r}\right)$ as the superposition
of a known incident wave function $\psi_{\mathrm{inc}}\left(\mathbf{r}\right)$
with energy $E$ and an additional wave function $\varPhi\left(\mathbf{r}\right)$
such that $\varPhi\left(\mathbf{r}\right)\rightarrow0$ for large
$\mathbf{r}$:
\begin{equation}
\psi\left(\mathbf{r}\right)=\begin{cases}
\psi_{1}\left(\mathbf{r}\right)\equiv\varPhi_{1}\left(\mathbf{r}\right) & \mathbf{r}\in\Omega_{1};\\
\psi_{2}\left(\mathbf{r}\right)\equiv\psi_{\mathrm{inc}}\left(\mathbf{r}\right)+\varPhi_{2}\left(\mathbf{r}\right) & \mathbf{r}\in\Omega_{2}.
\end{cases}\label{eq:newWavefunctionDefinition}
\end{equation}
Under these assumptions, (\ref{eq:system}) becomes: 
\begin{equation}
\Delta\varPhi_{j}\left(\mathbf{r}\right)+k_{j}^{2}\varPhi_{j}\left(\mathbf{r}\right)=0\quad j=1,2.
\end{equation}
Combining this equation with (\ref{eq:greenEqu}) and repeating the
procedure of Section \ref{subsec:IntegralEquations}, we obtain:
\begin{equation}
{\displaystyle \begin{cases}
\varPhi_{1}\left(\mathbf{r}\right)=\int_{S}d\mathbf{r}'\left[g_{1}\frac{\partial\varPhi_{1}\left(\mathbf{r}'\right)}{\partial n'}-\varPhi_{1}\left(\mathbf{r}'\right)\frac{\partial g_{1}}{\partial n'}\right] & \mathbf{r}\in\Omega_{1};\\
\varPhi_{2}\left(\mathbf{r}\right)=\int_{S}d\mathbf{r}'\left[\varPhi_{2}\left(\mathbf{r}'\right)\frac{\partial g_{2}}{\partial n'}-g_{2}\frac{\partial\varPhi_{2}\left(\mathbf{r}'\right)}{\partial n'}\right] & \mathbf{r}\in\Omega_{2}.
\end{cases}}\label{eq:scattSolSystem}
\end{equation}
Furthermore, by carefully taking the limit to the boundary:
\begin{equation}
{\displaystyle \begin{cases}
\frac{\varPhi_{1}\left(\mathbf{r}\right)}{2}=\fint_{S}d\mathbf{r}'\left[g_{1}\frac{\partial\varPhi_{1}\left(\mathbf{r}'\right)}{\partial n'}-\varPhi_{1}\left(\mathbf{r}'\right)\frac{\partial g_{1}}{\partial n'}\right] & \mathbf{r}\in S;\\
\frac{\varPhi_{2}\left(\mathbf{r}\right)}{2}=\fint_{S}d\mathbf{r}'\left[\varPhi_{2}\left(\mathbf{r}'\right)\frac{\partial g_{2}}{\partial n'}-g_{2}\frac{\partial\varPhi_{2}\left(\mathbf{r}'\right)}{\partial n'}\right] & \mathbf{r}\in S.
\end{cases}}\label{eq:scattSolSystemRegularized}
\end{equation}
Let now:
\begin{equation}
\varUpsilon_{j}\left(\mathbf{r}_{S}\right)\equiv\frac{1}{m_{j}}\frac{\partial\psi_{j}\left(\mathbf{r}_{S}\right)}{\partial n},\quad\chi_{j}\left(\mathbf{r}_{S}\right)\equiv\frac{1}{m_{j}}\frac{\partial\varPhi_{j}\left(\mathbf{r}_{S}\right)}{\partial n}\quad j=1,2.\label{eq:newNormalDerivatives}
\end{equation}
With this conventions, and taking into account (\ref{eq:boundaryLimits}),
the boundary conditions read:
\begin{equation}
{\displaystyle \begin{cases}
\delta\psi\left(\mathbf{r}_{S}\right)\equiv\psi_{2}\left(\mathbf{r}_{S}\right)-\psi_{1}\left(\mathbf{r}_{S}\right)=0\quad\Rightarrow & \psi_{2}\left(\mathbf{r}_{S}\right)=\psi_{1}\left(\mathbf{r}_{S}\right)=\varPhi_{1}\left(\mathbf{r}_{S}\right);\\
\delta\varUpsilon\left(\mathbf{r}_{S}\right)\equiv\varUpsilon_{2}\left(\mathbf{r}_{S}\right)-\varUpsilon_{1}\left(\mathbf{r}_{S}\right)=0\quad\Rightarrow & \varUpsilon_{2}\left(\mathbf{r}_{S}\right)=\varUpsilon_{1}\left(\mathbf{r}_{S}\right)=\chi_{1}\left(\mathbf{r}_{S}\right),
\end{cases}}\label{eq:newBoundaryConditionSystem}
\end{equation}
therefore:
\begin{equation}
\varPhi_{2}\left(\mathbf{r}_{S}\right)=\varPhi_{1}\left(\mathbf{r}_{S}\right)-\psi_{\mathrm{inc}}\left(\mathbf{r}_{S}\right),\quad\chi_{2}\left(\mathbf{r}_{S}\right)=\chi_{1}\left(\mathbf{r}_{S}\right)-\frac{1}{m_{2}}\frac{\partial\psi_{\mathrm{inc}}\left(\mathbf{r}_{S}\right)}{\partial n}.\label{eq:boundaryConditions}
\end{equation}
By redefining $\varPhi_{1}\left(\mathbf{r}_{S}\right)\equiv\varPhi\left(\mathbf{r}_{S}\right)$
and $\chi_{1}\left(\mathbf{r}_{S}\right)\equiv\chi\left(\mathbf{r}_{S}\right)$,
the first equation in system (\ref{eq:scattSolSystemRegularized})
is easily recast into the form:
\begin{equation}
\fint_{S}d\mathbf{r}_{S}'\left[\frac{\partial g_{1}\left(\mathbf{r}_{S},\mathbf{r}_{S}'\right)}{\partial n'}\varPhi\left(\mathbf{r}_{S}'\right)-m_{1}g_{1}\left(\mathbf{r}_{S},\mathbf{r}_{S}'\right)\chi\left(\mathbf{r}_{S}'\right)\right]=-\frac{\varPhi\left(\mathbf{r}_{S}\right)}{2}.\label{eq:systemEq1}
\end{equation}

We now combine the Helmholtz equation for $\psi_{\mathrm{inc}}\left(\mathbf{r}\right)$
with (\ref{eq:greenEqu}) to get:
\begin{equation}
g_{2}\left(\mathbf{r},\mathbf{r}'\right)\Delta'\psi_{\mathrm{inc}}\left(\mathbf{r}'\right)-\psi_{\mathrm{inc}}\left(\mathbf{r}'\right)\Delta'g_{2}\left(\mathbf{r},\mathbf{r}'\right)=\psi_{\mathrm{inc}}\left(\mathbf{r}'\right)\delta\left(\mathbf{r}-\mathbf{r}'\right).
\end{equation}
For $\mathbf{r}\in\Omega_{2}$, the integral of the above expression
over $\Omega_{1}$ can be rewritten using (\ref{eq:GreenId}):
\begin{align}
 & \int_{S}d\mathbf{r}_{S}'\left[g_{2}\left(\mathbf{r},\mathbf{r}_{S}'\right)\frac{\partial\psi_{\mathrm{inc}}\left(\mathbf{r}_{S}'\right)}{\partial n'}-\frac{\partial g_{2}\left(\mathbf{r},\mathbf{r}_{S}'\right)}{\partial n'}\psi_{\mathrm{inc}}\left(\mathbf{r}_{S}'\right)\right]=0\quad\mathbf{r}\in\Omega_{2}.\label{eq:greenForPsiIncAlternative}
\end{align}
Taking with due care the limit $\mathbf{r}\rightarrow\mathbf{r}_{S}$,
we obtain: 
\begin{align}
 & \fint_{S}d\mathbf{r}_{S}'\left[\frac{\partial g_{2}\left(\mathbf{r}_{S},\mathbf{r}_{S}'\right)}{\partial n'}\psi_{\mathrm{inc}}\left(\mathbf{r}_{S}'\right)-g_{2}\left(\mathbf{r}_{S},\mathbf{r}_{S}'\right)\frac{\partial\psi_{\mathrm{inc}}\left(\mathbf{r}_{S}'\right)}{\partial n'}\right]=-\frac{\psi_{\mathrm{inc}}\left(\mathbf{r}_{S}\right)}{2}
\end{align}
Then, considering the second equation in (\ref{eq:scattSolSystemRegularized}):
\begin{align}
 & \fint_{S}d\mathbf{r}_{S}'\left[\frac{\partial g_{2}\left(\mathbf{r}_{S},\mathbf{r}_{S}'\right)}{\partial n'}\varPhi_{2}\left(\mathbf{r}_{S}'\right)-g_{2}\left(\mathbf{r}_{S},\mathbf{r}_{S}'\right)\frac{\partial\varPhi_{2}\left(\mathbf{r}_{S}'\right)}{\partial n'}\right]=\frac{\varPhi_{2}\left(\mathbf{r}_{S}\right)}{2}
\end{align}
and summing the last two expressions, with reference to (\ref{eq:newWavefunctionDefinition})
and (\ref{eq:newNormalDerivatives}), leads to:
\begin{align}
 & \fint_{S}d\mathbf{r}_{S}'\left[\frac{\partial g_{2}\left(\mathbf{r}_{S},\mathbf{r}_{S}'\right)}{\partial n'}\psi_{2}\left(\mathbf{r}_{S}'\right)-m_{2}g_{2}\left(\mathbf{r}_{S},\mathbf{r}_{S}'\right)\varUpsilon_{2}\left(\mathbf{r}_{S}'\right)\right]=\frac{\psi_{2}\left(\mathbf{r}_{S}\right)}{2}-\psi_{\mathrm{inc}}\left(\mathbf{r}_{S}\right)
\end{align}
If we resort to (\ref{eq:newBoundaryConditionSystem}), $\varPhi_{1}\left(\mathbf{r}_{S}\right)\equiv\varPhi\left(\mathbf{r}_{S}\right)$
and $\chi_{1}\left(\mathbf{r}_{S}\right)\equiv\chi\left(\mathbf{r}_{S}\right)$,
the previous equation becomes:
\begin{align}
 & \fint_{S}d\mathbf{r}_{S}'\left[\frac{\partial g_{2}\left(\mathbf{r}_{S},\mathbf{r}_{S}'\right)}{\partial n'}\varPhi\left(\mathbf{r}_{S}'\right)-m_{2}g_{2}\left(\mathbf{r}_{S},\mathbf{r}_{S}'\right)\chi\left(\mathbf{r}_{S}'\right)\right]=\frac{\varPhi\left(\mathbf{r}_{S}\right)}{2}-\psi_{\mathrm{inc}}\left(\mathbf{r}_{S}\right).\label{eq:systemEq2Alternative}
\end{align}
Finally, the sum of (\ref{eq:systemEq2Alternative}) and (\ref{eq:systemEq1})
gives: 
\begin{equation}
\fint_{S}d\mathbf{r}_{S}'\left[\frac{\partial g_{1}\left(\mathbf{r}_{S},\mathbf{r}_{S}'\right)}{\partial n'}+\frac{\partial g_{2}\left(\mathbf{r}_{S},\mathbf{r}_{S}'\right)}{\partial n'}\right]\varPhi\left(\mathbf{r}_{S}'\right)-\fint_{S}d\mathbf{r}_{S}'\left[m_{1}g_{1}\left(\mathbf{r}_{S},\mathbf{r}_{S}'\right)+m_{2}g_{2}\left(\mathbf{r}_{S},\mathbf{r}_{S}'\right)\right]\chi\left(\mathbf{r}_{S}'\right)=-\psi_{\mathrm{inc}}\left(\mathbf{r}_{S}\right)\label{eq:firstEquAlternative}
\end{equation}
which constitutes the first integral equation of the BEM system. 

In order to arrive at the second equation, we first need to consider
(\ref{eq:parallelSurfacesFamily}) and compute the normal derivative
of (\ref{eq:scattSolSystem}) at $S_{\varepsilon}^{\mp}$, respectively:
\begin{equation}
{\displaystyle \begin{cases}
\frac{\partial\varPhi_{1}\left(\mathbf{r}\right)}{\partial n_{-}}=\int_{S}d\mathbf{r}'\left[\frac{\partial g_{1}}{\partial n_{-}}\frac{\partial\varPhi_{1}\left(\mathbf{r}'\right)}{\partial n'}-\varPhi_{1}\left(\mathbf{r}'\right)\frac{\partial^{2}g_{1}}{\partial n_{-}\partial n'}\right] & \mathbf{r}\in S_{\varepsilon}^{-};\\
\frac{\partial\varPhi_{2}\left(\mathbf{r}\right)}{\partial n_{+}}=\int_{S}d\mathbf{r}'\left[\varPhi_{2}\left(\mathbf{r}'\right)\frac{\partial^{2}g_{2}}{\partial n_{+}\partial n'}-\frac{\partial g_{2}}{\partial n_{+}}\frac{\partial\varPhi_{2}\left(\mathbf{r}'\right)}{\partial n'}\right] & \mathbf{r}\in S_{\varepsilon}^{+}.
\end{cases}}\label{eq:scattSystemNormDer}
\end{equation}
We also evaluate the normal derivative of (\ref{eq:greenForPsiIncAlternative})
at $S_{\varepsilon}^{+}$:
\begin{align}
 & \int_{S}d\mathbf{r}_{S}'\left[\frac{\partial^{2}g_{2}\left(\mathbf{r},\mathbf{r}_{S}'\right)}{\partial n_{+}\partial n'}\psi_{\mathrm{inc}}\left(\mathbf{r}_{S}'\right)-\frac{\partial g_{2}\left(\mathbf{r},\mathbf{r}_{S}'\right)}{\partial n_{+}}\frac{\partial\psi_{\mathrm{inc}}\left(\mathbf{r}_{S}'\right)}{\partial n'}\right]=0\quad\mathbf{r}\in S_{\varepsilon}^{+}.
\end{align}
Then, taking the second equation in (\ref{eq:scattSystemNormDer}):
\begin{align}
 & \int_{S}d\mathbf{r}_{S}'\left[\frac{\partial^{2}g_{2}\left(\mathbf{r},\mathbf{r}_{S}'\right)}{\partial n_{+}\partial n'}\varPhi_{2}\left(\mathbf{r}_{S}'\right)-\frac{\partial g_{2}\left(\mathbf{r},\mathbf{r}_{S}'\right)}{\partial n_{+}}\frac{\partial\varPhi_{2}\left(\mathbf{r}_{S}'\right)}{\partial n'}\right]=\frac{\partial\varPhi_{2}\left(\mathbf{r}\right)}{\partial n_{+}}\quad\mathbf{r}\in S_{\varepsilon}^{+}
\end{align}
and combining it with the above expression, we obtain:
\begin{align}
\frac{\partial\psi_{2}\left(\mathbf{r}\right)}{\partial n_{+}} & =\frac{\partial\psi_{\mathrm{inc}}\left(\mathbf{r}\right)}{\partial n_{+}}+\int_{S}d\mathbf{r}_{S}'\left[\frac{\partial^{2}g_{2}\left(\mathbf{r},\mathbf{r}_{S}'\right)}{\partial n_{+}\partial n'}\psi_{2}\left(\mathbf{r}_{S}'\right)-m_{2}\frac{\partial g_{2}\left(\mathbf{r},\mathbf{r}_{S}'\right)}{\partial n_{+}}\varUpsilon_{2}\left(\mathbf{r}_{S}'\right)\right]\quad\mathbf{r}\in S_{\varepsilon}^{+}.\label{eq:systNorDerAlternativeEq2}
\end{align}
On the other hand, we still have the first equation in (\ref{eq:scattSystemNormDer}):
\begin{equation}
\frac{\partial\varPhi_{1}\left(\mathbf{r}\right)}{\partial n_{-}}=\int_{S}d\mathbf{r}_{S}'\left[\frac{\partial g_{1}\left(\mathbf{r},\mathbf{r}_{S}'\right)}{\partial n_{-}}\frac{\partial\varPhi_{1}\left(\mathbf{r}_{S}'\right)}{\partial n'}-\varPhi_{1}\left(\mathbf{r}{}_{S}'\right)\frac{\partial^{2}g_{1}\left(\mathbf{r},\mathbf{r}_{S}'\right)}{\partial n_{-}\partial n'}\right]\quad\mathbf{r}\in S_{\varepsilon}^{-}.\label{eq:systNorDerAlternativeEq1}
\end{equation}
Dividing (\ref{eq:systNorDerAlternativeEq2}) by $m_{2}$, (\ref{eq:systNorDerAlternativeEq1})
by $m_{1}$ and taking the difference between the two under the limit
$\varepsilon\rightarrow0$ gives:
\begin{align}
&\int_{S}d\mathbf{r}_{S}'\left[\frac{1}{m_{1}}\frac{\partial^{2}g_{1}\left(\mathbf{r}_{S},\mathbf{r}_{S}'\right)}{\partial n\partial n'}+\frac{1}{m_{2}}\frac{\partial^{2}g_{2}\left(\mathbf{r}_{S},\mathbf{r}_{S}'\right)}{\partial n\partial n'}\right]\varPhi\left(\mathbf{r}_{S}'\right)-\fint_{S}d\mathbf{r}_{S}'\left[\frac{\partial g_{1}\left(\mathbf{r}_{S},\mathbf{r}_{S}'\right)}{\partial n}+\frac{\partial g_{2}\left(\mathbf{r}_{S},\mathbf{r}_{S}'\right)}{\partial n}\right]\chi\left(\mathbf{r}_{S}'\right)\nonumber\\
&=-\frac{1}{m_{2}}\frac{\partial\psi_{\mathrm{inc}}\left(\mathbf{r}_{S}\right)}{\partial n},\label{eq:secondEquAlternative}
\end{align}
which then completes our BEM system. Proceeding as in Section \ref{subsec:IntegralEquations},
equations (\ref{eq:firstEquAlternative}) and (\ref{eq:secondEquAlternative})
can be expressed in matrix form:
\begin{equation}
\hat{\mathbf{H}}\left[\begin{array}{c}
\varPhi\\
\chi
\end{array}\right]\left(\mathbf{r}_{S}\right)=\mathbf{J}\left(\mathbf{r}_{S}\right),\label{eq:matrixEquNew}
\end{equation}
where $\hat{\mathbf{H}}$ is the matrix integral operator defined
in (\ref{eq:operatorH}) and: 
\begin{equation}
\mathbf{J}\left(\mathbf{r}_{S}\right)\equiv\left(\begin{array}{c}
\psi_{\mathrm{inc}}\left(\mathbf{r}_{S}\right)\\
\frac{1}{m_{2}}\frac{\partial\psi_{\mathrm{inc}}\left(\mathbf{r}_{S}\right)}{\partial n}
\end{array}\right).
\end{equation}

Summarizing, the electron wave function in the two regions has been
rewritten through (\ref{eq:newWavefunctionDefinition}) in terms of
a known incident wave function $\psi_{\mathrm{inc}}\left(\mathbf{r}\right)$
and a scattered field $\varPhi\left(\mathbf{r}\right)$ which
satisfies:
\begin{equation}
{\displaystyle \begin{cases}
\varPhi_{1}\left(\mathbf{r}\right)=\int_{S}d\mathbf{r}_{S}'\left[m_{1}g_{1}\left(\mathbf{r},\mathbf{r}_{S}'\right)\chi\left(\mathbf{r}_{S}'\right)-\frac{\partial g_{1}\left(\mathbf{r},\mathbf{r}_{S}'\right)}{\partial n'}\varPhi\left(\mathbf{r}_{S}'\right)\right] & \mathbf{r}\in\Omega_{1};\\
\varPhi_{2}\left(\mathbf{r}\right)=\int_{S}d\mathbf{r}_{S}'\left[\frac{\partial g_{2}\left(\mathbf{r},\mathbf{r}_{S}'\right)}{\partial n'}\varPhi\left(\mathbf{r}_{S}'\right)-m_{2}g_{2}\left(\mathbf{r},\mathbf{r}_{S}'\right)\chi\left(\mathbf{r}_{S}'\right)\right] & \mathbf{r}\in\Omega_{2},
\end{cases}}\label{eq:BEMscatteringSolution}
\end{equation}
with $\varPhi\left(\mathbf{r}_{S}\right)$ and $\chi\left(\mathbf{r}_{S}\right)$
representing the solution of the matrix integral equation (\ref{eq:matrixEquNew}).
Contrary to \cite{Knipp1996}, the proposed formulation results in
a very concise form of the inhomogeneous term $\mathbf{J}\left(\mathbf{r}_{S}\right)$,
dictated only by the boundary restrictions of the incident wave function
and its normal derivative.

\subsection{Scattering amplitude}

When the incident wave function $\psi_{\mathrm{inc}}\left(\mathbf{r}\right)$
is taken to be a plane wave $\exp\left(i\mathbf{k}_{\mathrm{inc}}\cdot\mathbf{r}\right)$
with $\left|\mathbf{k}_{\mathrm{inc}}\right|=k_{2}$, in the outer
region at great distances from the boundary $S$ we have:
\begin{equation}
\psi_{2}\left(\mathbf{r}\right)\sim\exp\left(i\mathbf{k}_{\mathrm{inc}}\cdot\mathbf{r}\right)+g_{2}\left(\mathbf{r},0\right)F\left(\mathbf{k}\right),\label{eq:farField}
\end{equation}
where $\mathbf{k}\equiv k_{2}\hat{\mathbf{r}}$, 
\begin{equation}
g_{2}\left(\mathbf{r},0\right)\sim\begin{cases}
\frac{i}{4}\sqrt{\frac{2}{\pi k_{2}r}}\exp\left(ik_{2}r-i\frac{\pi}{4}\right) & \mathrm{2D};\\
\frac{1}{4\pi r}\exp\left(ik_{2}r\right) & \mathrm{3D}
\end{cases}
\end{equation}
and:
\begin{equation}
F\left(\mathbf{k}\right)\equiv\begin{cases}
2\sqrt{\pi k_{2}}\left(1-i\right)f^{(\mathrm{2D})}\left(\mathbf{k}\right) & \mathrm{2D};\\
4\pi f^{(\mathrm{3D})}\left(\mathbf{k}\right) & \mathrm{3D},
\end{cases}\label{eq:scatteringCrossSections}
\end{equation}
being $f^{(\mathrm{2D})}\left(\mathbf{k}\right)$ and $f^{(\mathrm{3D})}\left(\mathbf{k}\right)$
the differential scattering amplitudes in two and three dimensions,
respectively. Now, combining the second equation in (\ref{eq:scattSolSystem})
with (\ref{eq:greenForPsiIncAlternative}) for $\mathbf{r}\gg\mathbf{r}_{S}'$
and making use of the far-field approximation: 
\begin{equation}
k_{2}\left|\mathbf{r}-\mathbf{r}_{S}'\right|\sim k_{2}\left(r-\frac{\mathbf{r}\cdot\mathbf{r}_{S}'}{r}\right)=k_{2}r-\mathbf{k}\cdot\mathbf{r}_{S}',
\end{equation}
we can write:
\begin{equation}
\psi_{2}\left(\mathbf{r}\right)\sim\exp\left(i\mathbf{k}_{\mathrm{inc}}\cdot\mathbf{r}\right)-g_{2}\left(\mathbf{r},0\right)\int_{S}d\mathbf{r}_{S}'\left[i\mathbf{k}\cdot\mathbf{n}'\varPhi\left(\mathbf{r}_{S}'\right)+m_{2}\chi\left(\mathbf{r}_{S}'\right)\right]\exp\left(-i\mathbf{k}\cdot\mathbf{r}_{S}'\right).\label{eq:explicitFarField}
\end{equation}
From the comparison between (\ref{eq:farField}) and (\ref{eq:explicitFarField}),
it follows that:
\begin{equation}
F\left(\mathbf{k}\right)=-\int_{S}d\mathbf{r}_{S}'\left[i\mathbf{k}\cdot\mathbf{n}'\varPhi\left(\mathbf{r}_{S}'\right)+m_{2}\chi\left(\mathbf{r}_{S}'\right)\right]\exp\left(-i\mathbf{k}\cdot\mathbf{r}_{S}'\right).\label{eq:generalizedScatteringCrossSection}
\end{equation}

\subsection{Discretization of the operators}

The integral equations so far derived can be discretized just as in
Section \ref{subsec:Discretization}, after expanding the boundary
restrictions of the scattered field and of its inverse mass weighted
normal derivative on a set of node-based basis functions:
\begin{equation}
\varPhi\left(\mathbf{r}_{S}'\right)=\sum_{j}\alpha_{j}\,f_{j}\left(\mathbf{r}_{S}'\right);\quad\chi\left(\mathbf{r}_{S}'\right)=\mu^{-1}\sum_{j}\beta_{j}\,f_{j}\left(\mathbf{r}_{S}'\right).
\end{equation}
This results in the following Galerkin-discretized version of system
(\ref{eq:matrixEquNew}): 
\begin{equation}
\sum_{j}\mathbf{H}_{ij}\left[\begin{array}{c}
\alpha_{j}\\
\beta_{j}
\end{array}\right]=\mathbf{J}_{i},\label{eq:scattMatrixEquDiscrete}
\end{equation}
where $\mathbf{H}_{ij}$ is defined in (\ref{eq:operatorHdiscrete})
and:
\begin{equation}
\mathbf{J}_{i}\equiv\sum_{m\in i}\int_{S_{m}}d\mathbf{r}\left(\begin{array}{c}
\psi_{\mathrm{inc}}\left(\mathbf{r}\right)\\
\frac{\mu}{m_{2}}\frac{\partial\psi_{\mathrm{inc}}\left(\mathbf{r}\right)}{\partial n}
\end{array}\right)f_{i}^{m}\left(\mathbf{r}\right).
\end{equation}
A numerical solution to (\ref{eq:scattMatrixEquDiscrete}) is then
achieved by matrix inversion\footnote{See also \ref{sec:E}.}:
\begin{equation}
\left[\begin{array}{c}
\alpha_{i}\\
\beta_{i}
\end{array}\right]=\sum_{j}\mathbf{H}_{ij}^{-1}\mathbf{J}_{j}
\end{equation}
and makes it possible to determine the BEM wave function from (\ref{eq:BEMscatteringSolution}):
\begin{equation}
{\displaystyle \begin{cases}
\varPhi_{1}\left(\mathbf{r}\right)=\sum_{n}\int_{S_{n}}d\mathbf{r}'\sum_{j\in n}\left[\beta_{j}\frac{m_{1}}{\mu}g_{1}\left(\mathbf{r},\mathbf{r}'\right)-\alpha_{j}\frac{\partial g_{1}\left(\mathbf{r},\mathbf{r}'\right)}{\partial n'}\right]\,f_{j}^{n}\left(\mathbf{r}'\right) & \mathbf{r}\in\Omega_{1};\\
\varPhi_{2}\left(\mathbf{r}\right)=\sum_{n}\int_{S_{n}}d\mathbf{r}'\sum_{j\in n}\left[\alpha_{j}\frac{\partial g_{2}\left(\mathbf{r},\mathbf{r}'\right)}{\partial n'}-\beta_{j}\frac{m_{2}}{\mu}g_{2}\left(\mathbf{r},\mathbf{r}'\right)\right]\,f_{j}^{n}\left(\mathbf{r}'\right) & \mathbf{r}\in\Omega_{2},
\end{cases}}\label{eq:scattBEMwavefunction}
\end{equation}
as well as the BEM scattering amplitude from (\ref{eq:generalizedScatteringCrossSection}):
\begin{equation}
F\left(\mathbf{k}\right)=-\sum_{n}\int_{S_{n}}d\mathbf{r}'\sum_{j\in n}\left[i\alpha_{j}\mathbf{k}\cdot\mathbf{n}'+\beta_{j}\frac{m_{2}}{\mu}\right]\,f_{j}^{n}\left(\mathbf{r}'\right)\exp\left(-i\mathbf{k}\cdot\mathbf{r}'\right).\label{eq:scatteringCrossBEM}
\end{equation}

\subsection{Examples and comparisons}

\begin{figure}
\begin{centering}
\includegraphics[width=1\textwidth]{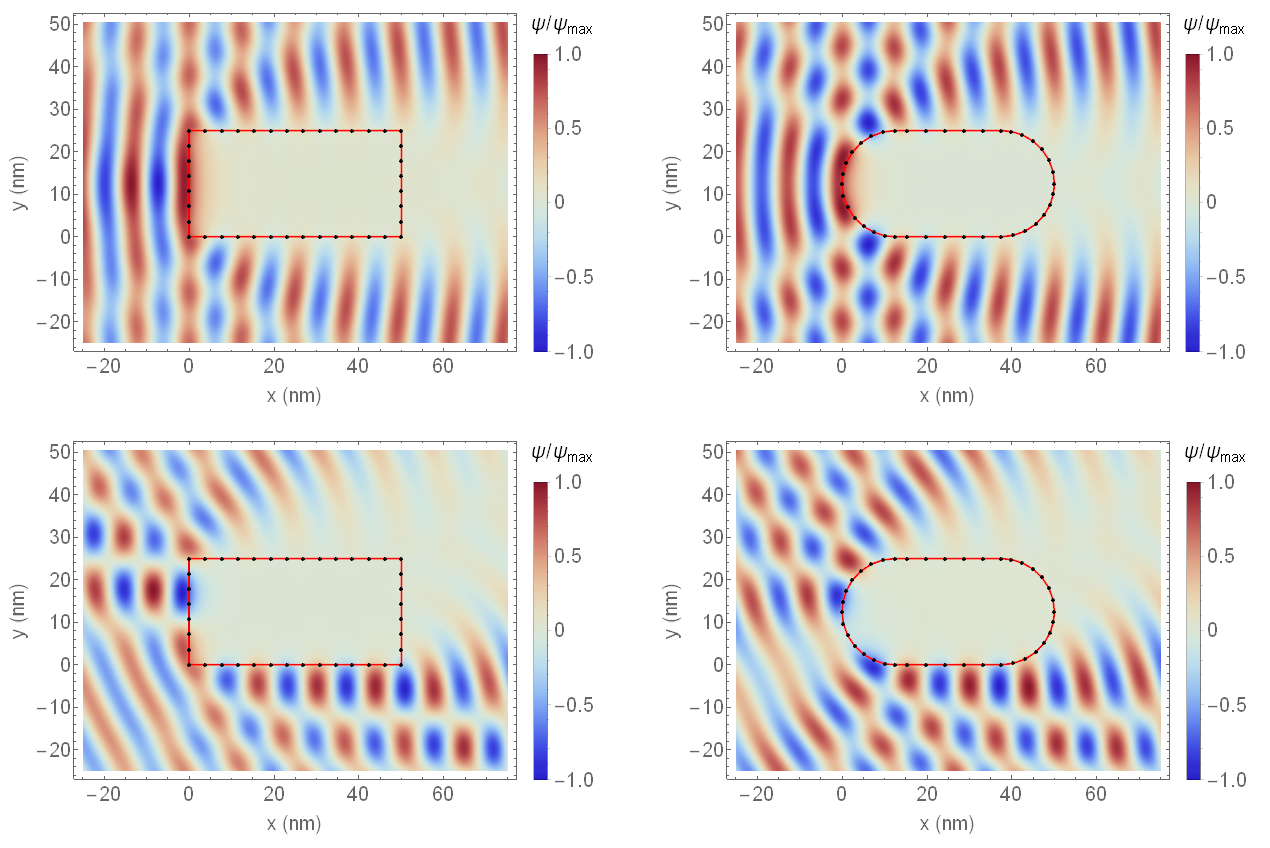}
\par\end{centering}
\caption{Scattering of an electron of energy $E=150\:\mathrm{meV}$ from a rectangular
(\textit{left}) and stadium-shaped (\textit{right}) quantum dot of
size $50\times25\:\mathrm{nm}^{2}$ with $m_{1}=m_{2}=0.0665\,m_{e}$
and $V=200\:\mathrm{meV}$. The total wave function is computed by the proposed
BEM formulation via (\ref{eq:scattMatrixEquDiscrete}) and (\ref{eq:scattBEMwavefunction})
for an electron impinging with $\theta_{\mathrm{inc}}=0$ (\textit{top}) and
with $\theta_{\mathrm{inc}}=\arctan\left(L_{y}/L_{x}\right)$ (\textit{bottom}),
using a mesh of $40$ elements, first-order basis functions and a
$10$ points Gauss-Legendre quadrature for the numerical integrations.\label{fig:scatteringStates}}
\end{figure}

\begin{figure}
\begin{centering}
\includegraphics[width=1\textwidth]{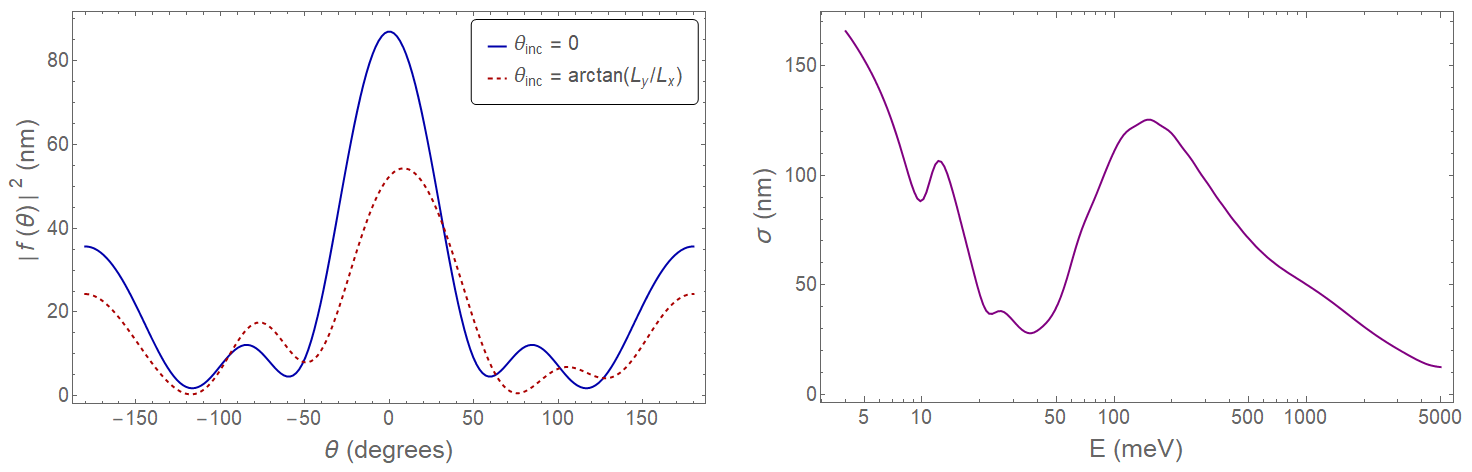}
\par\end{centering}
\caption{Scattering of an electron that moves along the $x$-axis from a rectangular
quantum dot of size $48\times24\:\mathrm{nm}^{2}$ with $m_{1}=m_{2}=0.0665\,m_{e}$
and $V=-50\:\mathrm{meV}$. \textit{Left}: differential scattering cross section
with respect to the $\theta$ angle for $E=5\:\mathrm{meV}$. \textit{Right}:
total scattering cross section (\ref{eq:crossSection2d}) as a function
of the electron energy. Both results are obtained by the proposed
BEM formulation via (\ref{eq:scattMatrixEquDiscrete}) and (\ref{eq:scatteringCrossBEM})
using a mesh of $40$ elements, first-order basis functions and a
$10$ points Gauss-Legendre quadrature for the numerical integrations.\label{fig:scatteringCrossSection}}
\end{figure}

Assuming $\psi_{\mathrm{inc}}\left(\mathbf{r}\right)=\exp\left(i\mathbf{k}_{\mathrm{inc}}\cdot\mathbf{r}\right)$
with: 
\begin{equation}
\mathbf{k}_{\mathrm{inc}}=k_{2}\cos\theta_{\mathrm{inc}}\,\hat{\mathbf{x}}+k_{2}\sin\theta_{\mathrm{inc}}\,\hat{\mathbf{y}},
\end{equation}
we now reintroduce the rectangle and stadium geometries considered
in Section \ref{subsec:boundExamples} and set the band offset $V$ in (\ref{eq:scattPotentialAndMass})
and the electron energy $E$ 
to be $200$ meV and $150$ meV, respectively. Figure \ref{fig:scatteringStates}
displays the total electron wave function (\ref{eq:newWavefunctionDefinition})
in the two geometries for two different values of $\theta_{\mathrm{inc}}$.
By choosing: 
\begin{equation}
\mathbf{k}=k_{2}\cos\theta\,\hat{\mathbf{x}}+k_{2}\sin\theta\,\hat{\mathbf{y}},\label{eq:k}
\end{equation}
the differential scattering amplitude $f^{(\mathrm{2D})}\left(\mathbf{k}\right)$
obtained from (\ref{eq:scatteringCrossSections}) and (\ref{eq:scatteringCrossBEM})
can be rewritten as a function of the angle $\theta$ and the total
scattering cross section is defined as:
\begin{equation}
\sigma\equiv\int_{0}^{2\pi}\left|f^{(\mathrm{2D})}\left(\theta\right)\right|^{2}d\theta.\label{eq:crossSection2d}
\end{equation}
For the sake of comparison, both $\left|f^{(\mathrm{2D})}\left(\theta\right)\right|^{2}$
and $\sigma$ are shown in Figure \ref{fig:scatteringCrossSection}
for a rectangular quantum dot like that considered in \cite{Knipp1996}.
As expected, the results match very well.

\section{Spectral density function\label{sec:Spectral}}

\subsection{Integral equations}

Taking $E_{\nu}$ and $\psi^{(\nu)}\left(\mathbf{r}\right)$ to
represent the energy and normalized wave function of the $\nu$-th
quantum state of the electron, the spectral density function:
\begin{equation}
\rho\left(\mathbf{r},\mathbf{r}';E\right)\equiv\sum_{\nu}\frac{\psi^{(\nu)}\left(\mathbf{r}\right)\left[\psi^{(\nu)}\left(\mathbf{r}'\right)\right]^{*}}{E_{\nu}-E}\label{eq:spectralDensity}
\end{equation}
provides a unified description of both the discrete and continuous
portions of the spectrum \cite{Knipp1996}. Within the arbitrary two-region
system so far considered, the spectral density function may be rewritten
as:
\begin{equation}
\rho\left(\mathbf{r},\mathbf{r}';E\right)\equiv\begin{cases}
\rho_{1}\left(\mathbf{r},\mathbf{r}';E\right) & \mathbf{r},\mathbf{r}'\in\Omega_{1};\\
\rho_{2}\left(\mathbf{r},\mathbf{r}';E\right) & \mathbf{r},\mathbf{r}'\in\Omega_{2}
\end{cases}
\end{equation}
and it is found to satisfy:
\begin{equation}
\Delta\rho_{j}\left(\mathbf{r},\mathbf{r}';E\right)+k_{j}^{2}\rho_{j}\left(\mathbf{r},\mathbf{r}';E\right)=-\frac{2m_{j}}{\hbar^{2}}\delta\left(\mathbf{r}-\mathbf{r}'\right)\quad j=1,2,\label{eq:spectralDensityEqu}
\end{equation}
as follows from (\ref{eq:compactEquation}) and from the completeness
relation:
\begin{equation}
\sum_{\nu}\psi^{(\nu)}\left(\mathbf{r}\right)\left[\psi^{(\nu)}\left(\mathbf{r}'\right)\right]^{*}=\delta\left(\mathbf{r}-\mathbf{r}'\right).
\end{equation}
Furthermore, by introducing:
\begin{equation}
\varPi_{j}\left(\mathbf{r}_{S},\mathbf{r}_{S}';E\right)\equiv\frac{1}{m_{j}}\frac{\partial\rho_{j}\left(\mathbf{r}_{S},\mathbf{r}_{S}';E\right)}{\partial n}\quad j=1,2,
\end{equation}
from (\ref{eq:boundaryCondSystem}) we have:
\begin{equation}
{\displaystyle \begin{cases}
\rho_{2}\left(\mathbf{r}_{S},\mathbf{r}_{S}';E\right)=\rho_{1}\left(\mathbf{r}_{S},\mathbf{r}_{S}';E\right)\equiv\rho\left(\mathbf{r}_{S},\mathbf{r}_{S}';E\right);\\
\varPi_{2}\left(\mathbf{r}_{S},\mathbf{r}_{S}';E\right)=\varPi_{1}\left(\mathbf{r}_{S},\mathbf{r}_{S}';E\right)\equiv\varPi\left(\mathbf{r}_{S},\mathbf{r}_{S}';E\right);
\end{cases}}\label{eq:spectralDensityBC}
\end{equation}

Equations (\ref{eq:greenEqu}) and (\ref{eq:spectralDensityEqu})
can be easily manipulated and combined to obtain:
\begin{equation}
g_{j}\left(\mathbf{r},\mathbf{r}'\right)\Delta'\rho_{j}\left(\mathbf{r}',\mathbf{r}''\right)-\rho_{j}\left(\mathbf{r}',\mathbf{r}''\right)\Delta'g_{j}\left(\mathbf{r},\mathbf{r}'\right)=\rho_{j}\left(\mathbf{r}',\mathbf{r}''\right)\delta\left(\mathbf{r}-\mathbf{r}'\right)-\frac{2m_{j}}{\hbar^{2}}g_{j}\left(\mathbf{r},\mathbf{r}'\right)\delta\left(\mathbf{r}'-\mathbf{r}''\right)\quad j=1,2,
\end{equation}
where $\mathbf{r},\,\mathbf{r}',\,\mathbf{r}''\in\Omega_{j}$
and the energy dependence of the spectral density function has been
suppressed for brevity. Integrating the expression in $d\mathbf{r}'$
over the volume $\Omega_{j}\setminus B_{\varrho}\left(\mathbf{r}\right)$
and using (\ref{eq:GreenId}), we get:
\begin{equation}
\int_{S}d\mathbf{r}_{S}'\left[g_{1}\left(\mathbf{r},\mathbf{r}_{S}'\right)\frac{\partial\rho_{1}\left(\mathbf{r}_{S}',\mathbf{r}''\right)}{\partial n'}-\rho_{1}\left(\mathbf{r}_{S}',\mathbf{r}''\right)\frac{\partial g_{1}\left(\mathbf{r},\mathbf{r}_{S}'\right)}{\partial n'}\right]=\rho_{1}\left(\mathbf{r},\mathbf{r}''\right)-\frac{2m_{1}}{\hbar^{2}}g_{1}\left(\mathbf{r},\mathbf{r}''\right)\label{eq:spectralDensityFirstEq}
\end{equation}
for $\mathbf{r},\,\mathbf{r}''\in\Omega_{1}$, and:
\begin{equation}
\int_{S}d\mathbf{r}_{S}'\left[\rho_{2}\left(\mathbf{r}_{S}',\mathbf{r}''\right)\frac{\partial g_{2}\left(\mathbf{r},\mathbf{r}_{S}'\right)}{\partial n'}-g_{2}\left(\mathbf{r},\mathbf{r}_{S}'\right)\frac{\partial\rho_{2}\left(\mathbf{r}_{S}',\mathbf{r}''\right)}{\partial n'}\right]=\rho_{2}\left(\mathbf{r},\mathbf{r}''\right)-\frac{2m_{2}}{\hbar^{2}}g_{2}\left(\mathbf{r},\mathbf{r}''\right)\label{eq:spectralDensitySecondEq}
\end{equation}
for $\mathbf{r},\,\mathbf{r}''\in\Omega_{2}$. Under the limits
$\mathbf{r}\rightarrow\mathbf{r}_{S}$ and $\mathbf{r}''\rightarrow\mathbf{r}_{S}''$,
(\ref{eq:spectralDensityFirstEq}) and (\ref{eq:spectralDensitySecondEq})
reduce to:
\begin{equation}
\fint_{S}d\mathbf{r}_{S}'\left[g_{1}\left(\mathbf{r}_{S},\mathbf{r}_{S}'\right)\frac{\partial\rho_{1}\left(\mathbf{r}_{S}',\mathbf{r}_{S}''\right)}{\partial n'}-\rho_{1}\left(\mathbf{r}_{S}',\mathbf{r}_{S}''\right)\frac{\partial g_{1}\left(\mathbf{r}_{S},\mathbf{r}_{S}'\right)}{\partial n'}\right]=\frac{\rho_{1}\left(\mathbf{r}_{S},\mathbf{r}_{S}''\right)}{2}-\frac{m_{1}}{\hbar^{2}}g_{1}\left(\mathbf{r}_{S},\mathbf{r}_{S}''\right)
\end{equation}
and:
\begin{equation}
\fint_{S}d\mathbf{r}_{S}'\left[\rho_{2}\left(\mathbf{r}_{S}',\mathbf{r}_{S}''\right)\frac{\partial g_{2}\left(\mathbf{r}_{S},\mathbf{r}_{S}'\right)}{\partial n'}-g_{2}\left(\mathbf{r}_{S},\mathbf{r}_{S}'\right)\frac{\partial\rho_{2}\left(\mathbf{r}_{S}',\mathbf{r}_{S}''\right)}{\partial n'}\right]=\frac{\rho_{2}\left(\mathbf{r}_{S},\mathbf{r}_{S}''\right)}{2}-\frac{m_{2}}{\hbar^{2}}g_{2}\left(\mathbf{r}_{S},\mathbf{r}_{S}''\right),
\end{equation}
respectively. Taking the difference between the two resulting expressions
with reference to (\ref{eq:spectralDensityBC}), we arrive at:
\begin{eqnarray}
&  & \fint_{S}d\mathbf{r}_{S}'\left[\frac{\partial g_{1}\left(\mathbf{r}_{S},\mathbf{r}_{S}'\right)}{\partial n'}+\frac{\partial g_{2}\left(\mathbf{r}_{S},\mathbf{r}_{S}'\right)}{\partial n'}\right]\rho\left(\mathbf{r}_{S}',\mathbf{r}_{S}''\right)-\fint_{S}d\mathbf{r}_{S}'\left[m_{1}g_{1}\left(\mathbf{r}_{S},\mathbf{r}_{S}'\right)+m_{2}g_{2}\left(\mathbf{r}_{S},\mathbf{r}_{S}'\right)\right]\varPi\left(\mathbf{r}_{S}',\mathbf{r}_{S}''\right)\nonumber\\ 
&  & =\frac{m_{1}}{\hbar^{2}}g_{1}\left(\mathbf{r}_{S},\mathbf{r}_{S}''\right)-\frac{m_{2}}{\hbar^{2}}g_{2}\left(\mathbf{r}_{S},\mathbf{r}_{S}''\right).\label{eq:spectral1}
\end{eqnarray}

As it is now customary, the second equation of the BEM system can
be obtained by combining the inverse mass weighted normal derivatives
of (\ref{eq:spectralDensityFirstEq}) and (\ref{eq:spectralDensitySecondEq})
at $\mathbf{r}\in S_{\varepsilon}^{\mp}$ under the limit $\varepsilon\rightarrow0$,
which leads to:
\begin{eqnarray}
&  &\int_{S}d\mathbf{r}_{S}'\left[\frac{1}{m_{1}}\frac{\partial^{2}g_{1}\left(\mathbf{r}_{S},\mathbf{r}_{S}'\right)}{\partial n\partial n'}+\frac{1}{m_{2}}\frac{\partial^{2}g_{2}\left(\mathbf{r}_{S},\mathbf{r}_{S}'\right)}{\partial n\partial n'}\right]\rho\left(\mathbf{r}_{S}',\mathbf{r}_{S}''\right)-\fint_{S}d\mathbf{r}_{S}'\left[\frac{\partial g_{1}\left(\mathbf{r}_{S},\mathbf{r}_{S}'\right)}{\partial n}+\frac{\partial g_{2}\left(\mathbf{r}_{S},\mathbf{r}_{S}'\right)}{\partial n}\right]\nonumber\\
&  & \times \varPi\left(\mathbf{r}_{S}',\mathbf{r}_{S}''\right) =\frac{1}{\hbar^{2}}\left[\frac{\partial g_{1}\left(\mathbf{r}_{S},\mathbf{r}_{S}''\right)}{\partial n}-\frac{\partial g_{2}\left(\mathbf{r}_{S},\mathbf{r}_{S}''\right)}{\partial n}\right].\label{eq:spectral2}
\end{eqnarray}
Equations (\ref{eq:spectral1}) and (\ref{eq:spectral2}) are rewritten
compactly as:
\begin{equation}
\hat{\mathbf{H}}\left[\begin{array}{c}
\rho\\
\varPi
\end{array}\right]\left(\mathbf{r}_{S},\mathbf{r}_{S}''\right)=\mathbf{G}\left(\mathbf{r}_{S},\mathbf{r}_{S}''\right),\label{eq:matrixEquSpectral}
\end{equation}
where $\hat{\mathbf{H}}$ is still the same as in (\ref{eq:operatorH})
and:
\begin{equation}
\mathbf{G}\left(\mathbf{r}_{S},\mathbf{r}_{S}''\right)\equiv\frac{1}{\hbar^{2}}\left(\begin{array}{c}
m_{2}g_{2}\left(\mathbf{r}_{S},\mathbf{r}_{S}''\right)-m_{1}g_{1}\left(\mathbf{r}_{S},\mathbf{r}_{S}''\right)\\
\frac{\partial g_{2}\left(\mathbf{r}_{S},\mathbf{r}_{S}''\right)}{\partial n}-\frac{\partial g_{1}\left(\mathbf{r}_{S},\mathbf{r}_{S}''\right)}{\partial n}
\end{array}\right).
\end{equation}
Finally, as suggested in \cite{Knipp1996}, the boundary data can
be condensed into the following distribution:
\begin{equation}
\rho_{w}\left(E\right)\equiv\int_{S}d\mathbf{r}_{S}\int_{S}d\mathbf{r}_{S}'\,w\left(\mathbf{r}_{S},\mathbf{r}_{S}'\right)\mathrm{Im}\left[\rho\left(\mathbf{r}_{S},\mathbf{r}_{S}';E\right)\right],\label{eq:condensedSpectralDensity}
\end{equation}
being $w\left(\mathbf{r}_{S},\mathbf{r}_{S}'\right)$ an arbitrary
weighting function.

\subsection{Discretization of the operators}

Since the boundary restrictions in (\ref{eq:matrixEquSpectral}) are
now functions of two space variables besides the electron energy,
their expansion on the set of node-based basis functions may be expressed
as:
\begin{equation}
\rho\left(\mathbf{r}_{S}',\mathbf{r}_{S}'';E\right)=\sum_{j,k}\alpha_{jk}\left(E\right)\,f_{j}\left(\mathbf{r}_{S}'\right)f_{k}\left(\mathbf{r}_{S}''\right);\quad\varPi\left(\mathbf{r}_{S}',\mathbf{r}_{S}'';E\right)=\mu^{-1}\sum_{j,k}\beta_{jk}\left(E\right)\,f_{j}\left(\mathbf{r}_{S}'\right)f_{k}\left(\mathbf{r}_{S}''\right).\label{eq:bfExpansionSpectral}
\end{equation}
Multiplying equation (\ref{eq:matrixEquSpectral}) by $f_{i}\left(\mathbf{r}_{S}\right)\,f_{l}\left(\mathbf{r}_{S}''\right)$
and integrating twice over $S$ using (\ref{eq:bfRestriction}), we
obtain:
\begin{equation}
\sum_{j,k}\mathbf{H}_{ij}\left[\begin{array}{c}
\alpha_{jk}\,F_{kl}\\
\beta_{jk}\,F_{kl}
\end{array}\right]=\mathbf{G}_{il},\label{eq:matrixEquSpectralDiscrete}
\end{equation}
where $\mathbf{H}_{ij}$ is defined in (\ref{eq:operatorHdiscrete}),
\begin{equation}
F_{kl}\equiv\sum_{c\in k\land l}\int_{S_{c}}d\mathbf{r}''\,f_{k}^{c}\left(\mathbf{r}''\right)f_{l}^{c}\left(\mathbf{r}''\right)\label{eq:F}
\end{equation}
identifies a sparse symmetric matrix\footnote{For instance, in the two-dimensional case with first-order basis functions,
it is easy to show that $F_{kl}=\left(L_{1}+L_{2}\right)/3$ for $k=l$,
$F_{kl}=L/6$ when the nodes $k$, $l$ are first neighbors and $F_{kl}=0$
otherwise, being $L$, $L_{1}$ and $L_{2}$ the sizes of the common
segments $S_{c}$.} whose only non-vanishing entries are those for which the mesh nodes
$k$ and $l$ belong to the same simplex $S_{c}$, and:
\begin{equation}
\mathbf{G}_{il}\equiv\frac{1}{\hbar^{2}}\sum_{m\in i}\,\sum_{p\in l}\int_{S_{m}}d\mathbf{r}\fint_{S_{p}}d\mathbf{r}''\left(\begin{array}{c}
m_{2}g_{2}\left(\mathbf{r},\mathbf{r}''\right)-m_{1}g_{1}\left(\mathbf{r},\mathbf{r}''\right)\\
\mu\left[\frac{\partial g_{2}\left(\mathbf{r},\mathbf{r}''\right)}{\partial n}-\frac{\partial g_{1}\left(\mathbf{r},\mathbf{r}''\right)}{\partial n}\right]
\end{array}\right)f_{i}^{m}\left(\mathbf{r}\right)f_{l}^{p}\left(\mathbf{r}''\right).
\end{equation}
From (\ref{eq:condensedSpectralDensity}) and (\ref{eq:bfExpansionSpectral}),
it follows that:
\begin{equation}
\rho_{w}\left(E\right)=\sum_{m,n}\int_{S_{m}}d\mathbf{r}\int_{S_{n}}d\mathbf{r}'\,w\left(\mathbf{r},\mathbf{r}'\right)\sum_{i\in m}\,\sum_{j\in n}\mathrm{Im}\left[\alpha_{ij}\left(E\right)\right]f_{i}^{m}\left(\mathbf{r}\right)f_{j}^{n}\left(\mathbf{r}'\right).
\end{equation}
Taking the function $w\left(\mathbf{r},\mathbf{r}'\right)$
to be the Dirac delta $\delta\left(\mathbf{r}-\mathbf{r}'\right)$,
the previous expression becomes:
\begin{equation}
\rho_{\delta}\left(E\right)=\sum_{i,j}\mathrm{Im}\left[\alpha_{ij}\left(E\right)\right]\sum_{c\in i\land j}\int_{S_{c}}d\mathbf{r}\,f_{i}^{m}\left(\mathbf{r}\right)f_{j}^{m}\left(\mathbf{r}\right)=\sum_{i,j}\mathrm{Im}\left[\alpha_{ij}\left(E\right)\right]F_{ij},\label{eq:rhoDelta}
\end{equation}
where the matrix $F$ has already been defined in (\ref{eq:F}). Then,
letting:
\begin{equation}
\left[\begin{array}{c}
\widetilde{\alpha}_{il}\\
\widetilde{\beta}_{il}
\end{array}\right]\equiv\sum_{k}\left[\begin{array}{c}
\alpha_{ik}\,F_{kl}\\
\beta_{ik}\,F_{kl}
\end{array}\right],\label{eq:alphaBetaTilde}
\end{equation}
equation (\ref{eq:matrixEquSpectralDiscrete}) can be directly inverted\footnote{See also \ref{sec:E}.}
to give:
\begin{equation}
\left[\begin{array}{c}
\widetilde{\alpha}_{il}\\
\widetilde{\beta}_{il}
\end{array}\right]=\sum_{j}\mathbf{H}_{ij}^{-1}\mathbf{G}_{jl}.
\end{equation}
It now becomes apparent that the knowledge of the matrix $\widetilde{\alpha}$
enables us to easily determine the spectral density function from
(\ref{eq:rhoDelta}) and (\ref{eq:alphaBetaTilde}):
\begin{equation}
\rho_{\delta}\left(E\right)=\mathrm{Im}\left\{ \mathrm{tr}\left[\widetilde{\alpha}\left(E\right)\right]\right\} .\label{eq:condensedSpectralDensityBEM}
\end{equation}
The conciseness of this last result may be regarded as a further advantage
of the proposed BEM formulation.

\subsection{Examples and comparisons}

\begin{figure}
\begin{centering}
\includegraphics[width=1\textwidth]{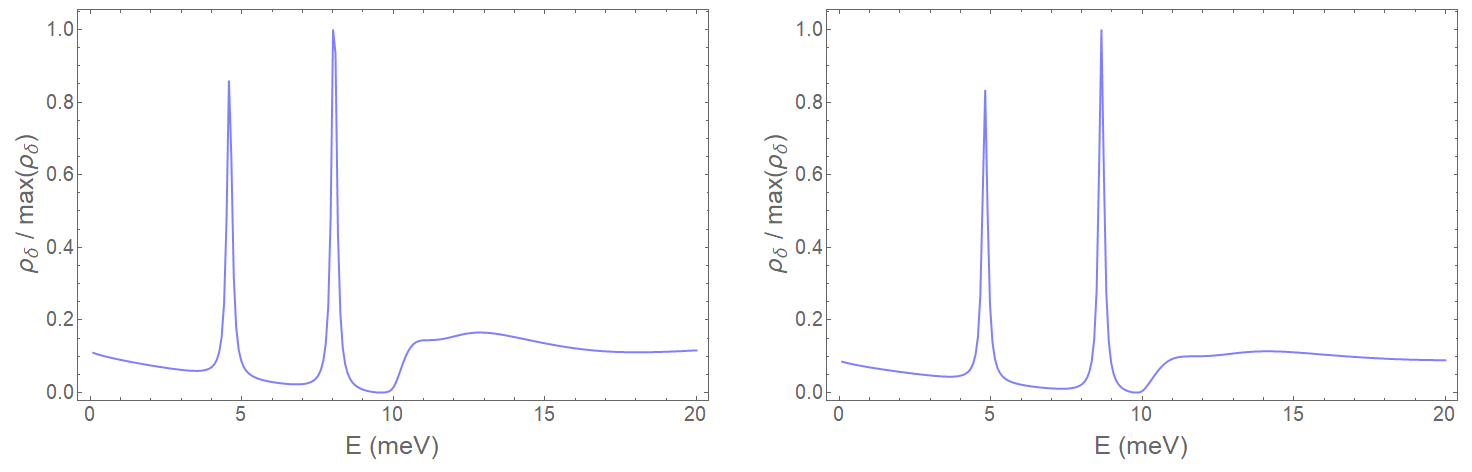}
\par\end{centering}
\caption{Spectral density of a rectangular (\textit{left}) and stadium-shaped
(\textit{right}) structure of size $50\times25\:\mathrm{nm}^{2}$
with $m_{1}=m_{2}=0.0665\,m_{e}$, $V=10\,\mathrm{meV}$ and $\mathrm{Im}\left(E\right)=0.1\:\mathrm{meV}$. The spectral density function is computed by the proposed BEM
formulation via (\ref{eq:matrixEquSpectralDiscrete}), (\ref{eq:alphaBetaTilde})
and (\ref{eq:condensedSpectralDensityBEM}) using a mesh of $40$
elements, first-order basis functions and a $10$ points Gauss-Legendre
quadrature for the numerical integrations.\label{fig:spectralDensity}}
\end{figure}

The BEM-computed spectral density function $\rho_{\delta}\left(E\right)$
is reported in Figure \ref{fig:spectralDensity} as a function of
the electron energy for both the previously considered rectangle and
stadium geometries.
 Since $\mathbf{H}$ becomes singular when $E$ approaches the
bound portion of the spectrum, analytic continuation to complex energies
may prove useful for display purposes, as explained in \cite{Knipp1996}.

\section{Conclusions}

As the examples throughout the paper testify, the proposed symmetric
Galerkin BEM gives very accurate results. Furthermore, it has the
advantage of leading to a simple implementation of the inhomogeneous
term in (\ref{eq:scattMatrixEquDiscrete}) and of the spectral density
function in (\ref{eq:condensedSpectralDensityBEM}). It is worth noting
that the integral equations (\ref{eq:matrixEqu}), (\ref{eq:matrixEquNew})
and (\ref{eq:matrixEquSpectral}) can be generalized to systems composed
of $N>2$ subregions. Most importantly, owing to the spectral properties
of the matrix integral operator (\ref{eq:operatorH}), both (\ref{eq:scattMatrixEquDiscrete})
and (\ref{eq:matrixEquSpectralDiscrete}) are suitable for preconditioning
strategies based on the Calderon identities, as detailed in
\ref{sec:E}. Despite direct inversion of the BEM matrix is not an
issue for the academic problems analyzed so far, preconditioned iterative
solvers may become essential to more realistic applications. The use
of fast algorithms to speed up the proposed BEM formulation will be
considered in future works.

\section*{Acknowledgments}
The authors wish to thank Prof. Francesco Andriulli for suggesting this study and two anonymous reviewers for their constructive comments which greatly improved the manuscript.  

\appendix

\section{Proof of formula (\ref{eq:sphericalIntegral})\label{sec:A}}

The free-space Green functions for the scalar Helmholtz equation in
one, two and three dimensions are defined, respectively, as follows:

\begin{align}
g^{(\mathrm{1D})}\left(x,x'\right) & \equiv\frac{i}{2k}\exp\left(ik\left|x-x'\right|\right);\\
g^{(\mathrm{2D})}\left(\boldsymbol{\rho},\boldsymbol{\rho}'\right) & \equiv\frac{i}{4}H_{0}^{(1)}\left(k\left|\boldsymbol{\rho}-\boldsymbol{\rho}'\right|\right);\label{eq:Green2D}\\
g^{(\mathrm{3D})}\left(\mathbf{r},\mathbf{r}'\right) & \equiv\frac{\exp\left(ik\left|\mathbf{r}-\mathbf{r}'\right|\right)}{4\pi\left|\mathbf{r}-\mathbf{r}'\right|}.
\end{align}
Let us write down the corresponding normal derivatives:
\begin{align}
\frac{\partial g^{(\mathrm{1D})}\left(x,x'\right)}{\partial n'} & =\frac{1}{2}\exp\left(ik\left|x-x'\right|\right)\mathrm{sgn}\left(x-x'\right)\frac{\partial x'}{\partial n'};\\
\frac{\partial g^{(\mathrm{2D})}\left(\boldsymbol{\rho},\boldsymbol{\rho}'\right)}{\partial n'} & =\frac{ikH_{1}^{(1)}\left(k\left|\boldsymbol{\rho}-\boldsymbol{\rho}'\right|\right)}{4\left|\boldsymbol{\rho}-\boldsymbol{\rho}'\right|}\left(\boldsymbol{\rho}-\boldsymbol{\rho}'\right)\cdot\mathbf{n}';\\
\frac{\partial g^{(\mathrm{3D})}\left(\mathbf{r},\mathbf{r}'\right)}{\partial n'} & =-\frac{\exp\left(ik\left|\mathbf{r}-\mathbf{r}'\right|\right)}{4\pi\left|\mathbf{r}-\mathbf{r}'\right|^{2}}\left(ik-\frac{1}{\left|\mathbf{r}-\mathbf{r}'\right|}\right)\left(\mathbf{r}-\mathbf{r}'\right)\cdot\mathbf{n}'.
\end{align}
In the one-dimensional case, we have:
\begin{align}
 & \lim_{\varrho\rightarrow0}\int_{\partial B_{\varrho}\left(\mathbf{r}\right)}d\mathbf{r}'\left[g^{(\mathrm{1D})}\frac{\partial\psi\left(\mathbf{r}'\right)}{\partial n'}-\psi\left(\mathbf{r}'\right)\frac{\partial g^{(\mathrm{1D})}}{\partial n'}\right]\nonumber \\
 & =\lim_{\varrho\rightarrow0}\left[g^{(\mathrm{1D})}\left(x,x+\varrho\right)\frac{\partial\psi\left(x+\varrho\right)}{\partial x}-g^{(\mathrm{1D})}\left(x,x-\varrho\right)\frac{\partial\psi\left(x-\varrho\right)}{\partial x}+\right.\nonumber \\
 & \left.-\psi\left(x+\varrho\right)\frac{\partial g^{(\mathrm{1D})}\left(x,x+\varrho\right)}{\partial n'}-\psi\left(x-\varrho\right)\frac{\partial g^{(\mathrm{1D})}\left(x,x-\varrho\right)}{\partial n'}\right]\nonumber \\
 & =\lim_{\varrho\rightarrow0}\left\{ \frac{i}{2k}\exp\left(ik\varrho\right)\left[\frac{\partial\psi\left(x+\varrho\right)}{\partial x}-\frac{\partial\psi\left(x-\varrho\right)}{\partial x}\right]+\frac{1}{2}\exp\left(ik\varrho\right)\left[\psi\left(x+\varrho\right)+\psi\left(x-\varrho\right)\right]\right\} =\psi\left(x\right).
\end{align}
Now, in two dimensions:
\begin{align}
 & \lim_{\varrho\rightarrow0}\int_{\partial B_{\varrho}\left(\mathbf{r}\right)}d\mathbf{r}'\left[g^{(\mathrm{2D})}\frac{\partial\psi\left(\mathbf{r}'\right)}{\partial n'}-\psi\left(\mathbf{r}'\right)\frac{\partial g^{(\mathrm{2D})}}{\partial n'}\right]\nonumber \\
 & =\lim_{\varrho\rightarrow0}\int_{0}^{2\pi}d\varphi\,\varrho\left[\frac{i}{4}H_{0}^{(1)}\left(k\varrho\right)\frac{\partial\psi\left(\boldsymbol{\rho}'\right)}{\partial n'}-\psi\left(\boldsymbol{\rho}'\right)\frac{ikH_{1}^{(1)}\left(k\varrho\right)}{4\varrho}\left(\boldsymbol{\rho}-\boldsymbol{\rho}'\right)\cdot\mathbf{n}'\right]\nonumber \\
 & =\lim_{\varrho\rightarrow0}\int_{0}^{2\pi}d\varphi\,\varrho\left\{ \frac{i}{4}\frac{\partial\psi\left(\boldsymbol{\rho}'\right)}{\partial n'}-\frac{1}{2\pi}\left[\log\left(\frac{k\varrho}{2}\right)+\gamma\right]\frac{\partial\psi\left(\boldsymbol{\rho}'\right)}{\partial n'}+\psi\left(\boldsymbol{\rho}'\right)\frac{\varrho}{2\pi\varrho^{2}}\right\} =\psi\left(\boldsymbol{\rho}\right),
\end{align}
where the following expansions of the Hankel functions for small argument
have been employed \cite{Abramowitz1972}:
\begin{align}
H_{0}^{(1)}\left(z\right) & \sim1+\frac{2i}{\pi}\left(\log z+\gamma-\log2\right);\\
H_{1}^{(1)}\left(z\right) & \sim-\frac{2i}{\pi z}.
\end{align}
The three-dimensional case is treated similarly:
\begin{align}
 & \lim_{\varrho\rightarrow0}\int_{\partial B_{\varrho}\left(\mathbf{r}\right)}d\mathbf{r}'\left[g^{(\mathrm{3D})}\frac{\partial\psi\left(\mathbf{r}'\right)}{\partial n'}-\psi\left(\mathbf{r}'\right)\frac{\partial g^{(\mathrm{3D})}}{\partial n'}\right]\nonumber \\
 & =\lim_{\varrho\rightarrow0}\int_{0}^{2\pi}d\phi\int_{0}^{\pi}d\theta\,\varrho^{2}\sin\theta\left[\frac{\partial\psi\left(\mathbf{r}'\right)}{\partial n'}-\psi\left(\mathbf{r}'\right)\left(\frac{ik}{\varrho}-\frac{1}{\varrho^{2}}\right)\varrho\right]\frac{\exp\left(ik\varrho\right)}{4\pi\varrho}=\psi\left(\mathbf{r}\right).
\end{align}

\section{Proof of formula (\ref{eq:SolSystemRegularized})\label{sec:B}}

\begin{figure}
\begin{centering}
\includegraphics[scale=0.3]{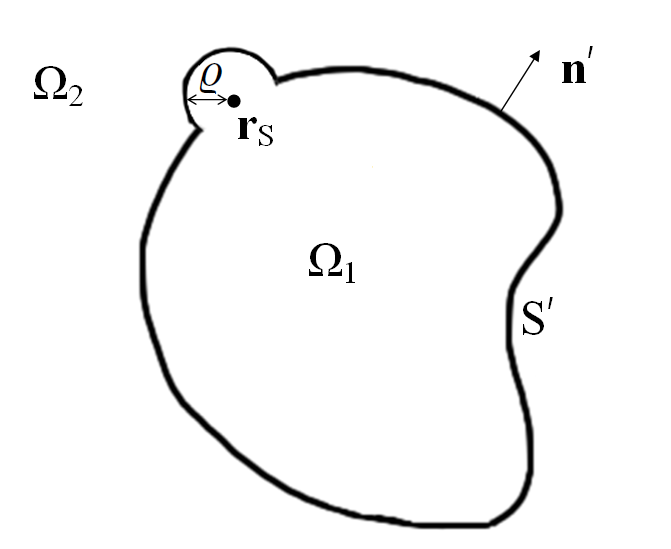}
\includegraphics[scale=0.3]{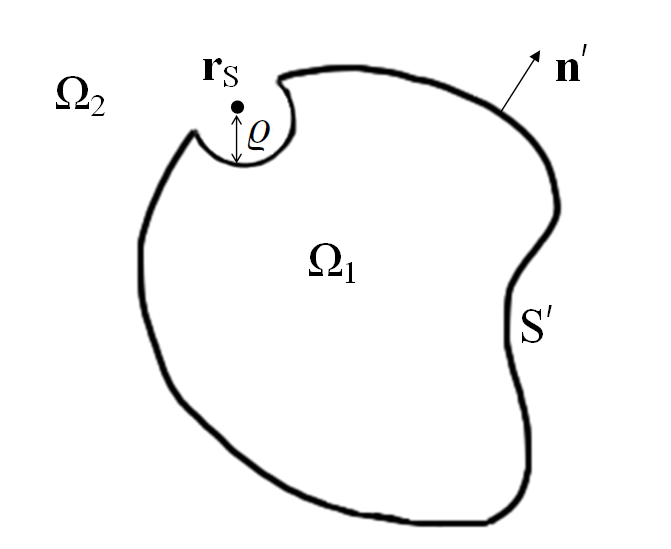}
\par\end{centering}
\caption{Sketch of two possible deformations of the integration surface.\label{fig:deformedSurf}}
\end{figure}

Before taking the limit $\mathbf{r}\rightarrow\mathbf{r}_{S}$,
let us assume to deform the integration surface in the first and second
equations of (\ref{eq:SolSystem}) as depicted in the left and right
sides of Figure \ref{fig:deformedSurf}, respectively, so that the
boundary integrals can be split into two parts: 
\begin{equation}
\int_{S}d\mathbf{r}'=\int_{S-\sigma_{\varrho}\left(\mathbf{r}_{S}\right)}d\mathbf{r}'+\int_{H\partial B_{\varrho}^{\pm}\left(\mathbf{r}_{S}\right)}d\mathbf{r}',\label{eq:splitIntegrals}
\end{equation}
being $\sigma_{\varrho}\left(\mathbf{r}_{S}\right)$ and $H\partial B_{\varrho}^{\pm}\left(\mathbf{r}_{S}\right)$
a disk and the upper and lower hemispheres of radius $\varrho$ centered
at $\mathbf{r}_{S}$.
 In the limit $\varrho\rightarrow0$, the first term in (\ref{eq:splitIntegrals})
may be replaced by a principal value integral. By writing explicitly
the second term in two dimensions, we get:
\begin{align}
 & \lim_{\varrho\rightarrow0}\int_{H\partial B_{\varrho}^{\pm}\left(\mathbf{r}_{S}\right)}d\mathbf{r}'\left[g^{(\mathrm{2D})}\left(\mathbf{r}_{S},\mathbf{r}'\right)\frac{\partial\psi\left(\mathbf{r}'\right)}{\partial n'}-\psi\left(\mathbf{r}'\right)\frac{\partial g^{(\mathrm{2D})}\left(\mathbf{r}_{S},\mathbf{r}'\right)}{\partial n'}\right]\nonumber \\
 & =\lim_{\varrho\rightarrow0}\int_{0}^{\pi}d\varphi\,\varrho\left[\frac{i}{4}H_{0}^{(1)}\left(k\varrho\right)\frac{\partial\psi\left(\boldsymbol{\rho}'\right)}{\partial n'}-\psi\left(\boldsymbol{\rho}'\right)\frac{ikH_{1}^{(1)}\left(k\varrho\right)}{4\varrho}\left(\mp\varrho\right)\right]=\pm\frac{\psi\left(\boldsymbol{\rho}_{S}\right)}{2}.
\end{align}
Analogously, in three dimensions:
\begin{align}
 & \lim_{\varrho\rightarrow0}\int_{H\partial B_{\varrho}^{\pm}\left(\mathbf{r}_{S}\right)}d\mathbf{r}'\left[g^{(\mathrm{3D})}\left(\mathbf{r}_{S},\mathbf{r}'\right)\frac{\partial\psi\left(\mathbf{r}'\right)}{\partial n'}-\psi\left(\mathbf{r}'\right)\frac{\partial g^{(\mathrm{3D})}\left(\mathbf{r}_{S},\mathbf{r}'\right)}{\partial n'}\right]\nonumber \\
 & =\lim_{\varrho\rightarrow0}\int_{0}^{2\pi}d\phi\int_{0}^{\pi/2}d\theta\,\varrho^{2}\sin\theta\left[\frac{\partial\psi\left(\mathbf{r}'\right)}{\partial n'}+\psi\left(\mathbf{r}'\right)\left(\frac{ik}{\varrho}-\frac{1}{\varrho^{2}}\right)\left(\mp\varrho\right)\right]\frac{\exp\left(ik\varrho\right)}{4\pi\varrho}=\pm\frac{\psi\left(\mathbf{r}_{S}\right)}{2}.
\end{align}

\section{Proof of formula (\ref{eq:secondEqu})\label{sec:C}}

By using $\psi_{1}\left(\mathbf{r}_{S}\right)\equiv\psi\left(\mathbf{r}_{S}\right)$
and $\chi_{1}\left(\mathbf{r}_{S}\right)\equiv\chi\left(\mathbf{r}_{S}\right)$,
system (\ref{eq:systemNormDer}) can be rewritten as: 
\begin{equation}
{\displaystyle \begin{cases}
\frac{1}{m_{1}}\frac{\partial\psi_{1}\left(\mathbf{r}_{-}\right)}{\partial n_{-}}=\int_{S}d\mathbf{r}_{S}'\frac{\partial g_{1}\left(\mathbf{r}_{-},\mathbf{r}_{S}'\right)}{\partial n_{-}}\chi\left(\mathbf{r}_{S}'\right)-\frac{1}{m_{1}}\int_{S}d\mathbf{r}_{S}'\frac{\partial^{2}g_{1}\left(\mathbf{r}_{-},\mathbf{r}_{S}'\right)}{\partial n_{-}\partial n'}\psi\left(\mathbf{r}_{S}'\right);\\
\frac{1}{m_{2}}\frac{\partial\psi_{2}\left(\mathbf{r}_{+}\right)}{\partial n_{+}}=\frac{1}{m_{2}}\int_{S}d\mathbf{r}_{S}'\frac{\partial^{2}g_{2}\left(\mathbf{r}_{+},\mathbf{r}_{S}'\right)}{\partial n_{+}\partial n'}\psi\left(\mathbf{r}_{S}'\right)-\int_{S}d\mathbf{r}_{S}'\frac{\partial g_{2}\left(\mathbf{r}_{+},\mathbf{r}_{S}'\right)}{\partial n_{+}}\chi\left(\mathbf{r}_{S}'\right),
\end{cases}}\label{eq:weightedSystem}
\end{equation}
where $\mathbf{r}_{\pm}\in S_{\varepsilon}^{\pm}$. The procedure
described in \ref{sec:B} is now applied to the integrals
containing a single normal derivative of the Green function, deforming
the integration surface and then taking the limit $\varepsilon\rightarrow0$,
so that $S_{\varepsilon}^{\pm}\rightarrow S$ and $\mathbf{r}_{\pm}\rightarrow\mathbf{r}_{S}$:
\begin{equation}
\int_{S}d\mathbf{r}_{S}'\frac{\partial g\left(\mathbf{r}_{S},\mathbf{r}_{S}'\right)}{\partial n}\chi\left(\mathbf{r}_{S}'\right)=\fint_{S}d\mathbf{r}_{S}'\frac{\partial g\left(\mathbf{r}_{S},\mathbf{r}_{S}'\right)}{\partial n}\chi\left(\mathbf{r}_{S}'\right)+\lim_{\varrho\rightarrow0}\int_{H\partial B_{\varrho}^{\pm}\left(\mathbf{r}_{S}\right)}d\mathbf{r}'\frac{\partial g\left(\mathbf{r}_{S},\mathbf{r}'\right)}{\partial n}\chi\left(\mathbf{r}'\right).
\end{equation}
Finally, making use of the result:
\begin{equation}
\lim_{\varrho\rightarrow0}\int_{H\partial B_{\varrho}^{+}\left(\mathbf{r}_{S}\right)}d\mathbf{r}'\frac{\partial g\left(\mathbf{r}_{S},\mathbf{r}'\right)}{\partial n}\chi\left(\mathbf{r}'\right)=-\lim_{\varrho\rightarrow0}\int_{H\partial B_{\varrho}^{-}\left(\mathbf{r}_{S}\right)}d\mathbf{r}'\frac{\partial g\left(\mathbf{r}_{S},\mathbf{r}'\right)}{\partial n}\chi\left(\mathbf{r}'\right),
\end{equation}
it is apparent that (\ref{eq:secondEqu}) corresponds to the difference
between the two equations in (\ref{eq:weightedSystem}).

\section{Semi-analytical formulas for the singular integrals in (\ref{eq:singleLayer})-(\ref{eq:hypersingular})\label{sec:D}}

Let us introduce the following boundary integral operators:
\begin{align}
\hat{\mathsf{s}}\left[f\right]\left(\mathbf{r}_{S}\right) & \equiv\fint_{S}d\mathbf{r}_{S}'\,g\left(\mathbf{r}_{S},\mathbf{r}_{S}'\right)f\left(\mathbf{r}_{S}'\right);\label{eq:SL}\\
\hat{\mathsf{d}}\left[f\right]\left(\mathbf{r}_{S}\right) & \equiv\fint_{S}d\mathbf{r}_{S}'\frac{\partial g\left(\mathbf{r}_{S},\mathbf{r}_{S}'\right)}{\partial n'}f\left(\mathbf{r}_{S}'\right);\\
\hat{\mathsf{d}}^{\dagger}\left[f\right]\left(\mathbf{r}_{S}\right) & \equiv\fint_{S}d\mathbf{r}_{S}'\frac{\partial g\left(\mathbf{r}_{S},\mathbf{r}_{S}'\right)}{\partial n}f\left(\mathbf{r}_{S}'\right);\\
\hat{\mathsf{n}}\left[f\right]\left(\mathbf{r}_{S}\right) & \equiv\int_{S}d\mathbf{r}_{S}'\frac{\partial^{2}g\left(\mathbf{r}_{S},\mathbf{r}_{S}'\right)}{\partial n\partial n'}f\left(\mathbf{r}_{S}'\right)\label{eq:HO}
\end{align}
and the corresponding Galerkin matrices:
\begin{align}
\mathsf{s}_{ij} & \equiv\sum_{m\in i}\,\sum_{n\in j}\int_{S_{m}}d\mathbf{r}\fint_{S_{n}}d\mathbf{r}'g\left(\mathbf{r},\mathbf{r}'\right)f_{i}^{m}\left(\mathbf{r}\right)f_{j}^{n}\left(\mathbf{r}'\right);\label{eq:SLdiscrete}\\
\mathsf{d}_{ij} & \equiv\sum_{m\in i}\,\sum_{n\in j}\int_{S_{m}}d\mathbf{r}\fint_{S_{n}}d\mathbf{r}'\frac{\partial g\left(\mathbf{r},\mathbf{r}'\right)}{\partial n'}f_{i}^{m}\left(\mathbf{r}\right)f_{j}^{n}\left(\mathbf{r}'\right);\label{eq:DLdiscrete}\\
\mathsf{d}_{ij}^{\dagger} & \equiv\sum_{m\in i}\,\sum_{n\in j}\int_{S_{m}}d\mathbf{r}\fint_{S_{n}}d\mathbf{r}'\frac{\partial g\left(\mathbf{r},\mathbf{r}'\right)}{\partial n}f_{i}^{m}\left(\mathbf{r}\right)f_{j}^{n}\left(\mathbf{r}'\right);\label{eq:ADLdiscrete}\\
\mathsf{n}_{ij} & \equiv\sum_{m\in i}\,\sum_{n\in j}\int_{S_{m}}d\mathbf{r}\int_{S_{n}}d\mathbf{r}'\frac{\partial^{2}g\left(\mathbf{r},\mathbf{r}'\right)}{\partial n\partial n'}f_{i}^{m}\left(\mathbf{r}\right)f_{j}^{n}\left(\mathbf{r}'\right).\label{eq:HOdiscrete}
\end{align}
Using these definitions, equations (\ref{eq:singleLayer})-(\ref{eq:hypersingular})
can be rewritten as:
\begin{align}
S & =m_{1}\mathsf{s}_{1}+m_{2}\mathsf{s}_{2};\label{eq:singleLayerExpression}\\
D & =\mathsf{d}_{1}+\mathsf{d}_{2};\\
D^{\dagger} & =\mathsf{d}_{1}^{\dagger}+\mathsf{d}_{2}^{\dagger};\\
N & =\frac{\mathsf{n}_{1}}{m_{1}}+\frac{\mathsf{n}_{2}}{m_{2}},\label{eq:hypersingularExpression}
\end{align}
where labels $1$, $2$ refer to the inner and outer regions, respectively,
and the matrix indices have been neglected to avoid confusion. 

Limiting the analysis to the two-dimensional case, where the piecewise
smooth closed curve $S$ is discretized into a collection of segments
$\left\{ S_{n}\right\} $ with lengths $\left\{ l_{n}\right\} $ and
extrema $\left\{ \mathbf{r}_{A}^{n};\:\mathbf{r}_{B}^{n}\right\} $,
we can consider first-order basis functions:
\begin{equation}
f_{j}^{n}\left(t_{n}\right)\equiv\begin{cases}
1-t_{n} & \mathrm{if}\:\mathbf{r}_{j}=\mathbf{r}_{A}^{n};\\
t_{n} & \mathrm{if}\:\mathbf{r}_{j}=\mathbf{r}_{B}^{n},
\end{cases}\label{eq:basisFunctions}
\end{equation}
being $\mathbf{r}_{j}=\left(x_{j},y_{j}\right)$ the position
of the $j$-th mesh node and $t_{n}\in\left[0,1\right]$ a local parameter
for the $n$-th segment such that:
\begin{equation}
\mathbf{r}\left(t_{n}\right)=\mathbf{r}_{A}^{n}+\left(\mathbf{r}_{B}^{n}-\mathbf{r}_{A}^{n}\right)t_{n}.\label{eq:parametricCoordinates}
\end{equation}
Each of the discrete operators $\mathsf{s}_{ij}$, $\mathsf{d}_{ij}$,
$\mathsf{d}_{ij}^{\dagger}$ and $\mathsf{n}_{ij}$ in (\ref{eq:SLdiscrete})-(\ref{eq:HOdiscrete})
consists of a sum of four double integrals over the pairs of segments
$\left(S_{m},S_{n}\right)\in\left\{ m\in i\right\} \times\left\{ n\in j\right\} $.
Whereas all the integrations involving disjoint segments can be computed
by Gauss-Legendre quadrature rules, the singular integrals over coincident
and adjacent segments deserve special care, as detailed in the following.

\subsection{Singular integrals in (\ref{eq:SLdiscrete})\label{subsec:D1}}

When $S_{m}=S_{n}$, the singular double integrals in (\ref{eq:SLdiscrete})
can be expressed as:
\begin{equation}
\frac{i}{4}\int_{0}^{l_{n}}dx\fint_{-x}^{l_{n}-x}dx'H_{0}^{(1)}\left(k\left|x'\right|\right)f_{i}^{n}\left(x\right)\widetilde{f}_{j}^{n}\left(x',x\right),\label{eq:singIntRewritten}
\end{equation}
with:
\begin{equation}
f_{i}^{n}\left(x\right)\equiv\begin{cases}
1-\frac{x}{l_{n}} & \mathrm{if}\:x_{i}=0;\\
\frac{x}{l_{n}} & \mathrm{if}\:x_{i}=l_{n},
\end{cases}\label{eq:bfi}
\end{equation}
\begin{equation}
\widetilde{f}_{j}^{n}\left(x',x\right)\equiv\begin{cases}
1-\frac{\left(x'+x\right)}{l_{n}} & \mathrm{if}\:x_{j}'=-x;\\
\frac{\left(x'+x\right)}{l_{n}} & \mathrm{if}\:x_{j}'=l_{n}-x.
\end{cases}\label{eq:bfj}
\end{equation}
Four kinds of integrals are obtained from (\ref{eq:singIntRewritten}),
(\ref{eq:bfi}) and (\ref{eq:bfj}), namely:
\begin{align}
I_{a}^{n} & \equiv\frac{i}{4}\int_{0}^{l_{n}}dx\fint_{-x}^{l_{n}-x}dx'H_{0}^{(1)}\left(k\left|x'\right|\right)\left(1-\frac{x}{l_{n}}\right)\left[1-\frac{\left(x'+x\right)}{l_{n}}\right];\label{eq:I11}\\
I_{b}^{n} & \equiv\frac{i}{4}\int_{0}^{l_{n}}dx\fint_{-x}^{l_{n}-x}dx'H_{0}^{(1)}\left(k\left|x'\right|\right)\left(1-\frac{x}{l_{n}}\right)\left[\frac{\left(x'+x\right)}{l_{n}}\right];\\
I_{c}^{n} & \equiv\frac{i}{4}\int_{0}^{l_{n}}dx\fint_{-x}^{l_{n}-x}dx'H_{0}^{(1)}\left(k\left|x'\right|\right)\left(\frac{x}{l_{n}}\right)\left[1-\frac{\left(x'+x\right)}{l_{n}}\right];\\
I_{d}^{n} & \equiv\frac{i}{4}\int_{0}^{l_{n}}dx\fint_{-x}^{l_{n}-x}dx'H_{0}^{(1)}\left(k\left|x'\right|\right)\left(\frac{x}{l_{n}}\right)\left[\frac{\left(x'+x\right)}{l_{n}}\right].\label{eq:I22}
\end{align}
Making reference to the tabulated formulas for Bessel functions \cite{Gradshteyn2007},
the previous expressions can be reduced to: 
\begin{align}
I_{a}^{n}=I_{d}^{n} & =\frac{i\pi}{8k^{2}}\left[H_{1}^{(1)}\left(kl_{n}\right)\mathbf{H}_{0}\left(kl_{n}\right)-H_{0}^{(1)}\left(kl_{n}\right)\mathbf{H}_{1}\left(kl_{n}\right)\right]+\nonumber \\
 & -\frac{i}{2k^{2}}H_{2}^{(1)}\left(kl_{n}\right)+\frac{2}{\pi k^{4}l_{n}^{2}}+\frac{i}{4k^{2}}\varGamma_{1}\left(kl_{n}\right)+\frac{i}{2k^{4}l_{n}^{2}}\varGamma_{3}\left(kl_{n}\right);\label{eq:I11I22}\\
I_{b}^{n}=I_{c}^{n} & =-I_{a}^{n}+\frac{i}{4k^{2}}\varGamma_{1}\left(kl_{n}\right),\label{eq:I12I21}
\end{align}
where $\mathbf{H}_{\nu}$ represents the Struve function of order
$\nu$ and the remaining integrals:
\begin{align}
\varGamma_{1}\left(\sigma\right) & \equiv\frac{\pi}{2}\int_{0}^{\sigma}dx\,x\left[H_{0}^{(1)}\left(x\right)\mathbf{H}_{-1}\left(x\right)+H_{1}^{(1)}\left(x\right)\mathbf{H}_{0}\left(x\right)\right];\label{eq:gamma0}\\
\varGamma_{3}\left(\sigma\right) & \equiv\frac{\pi}{2}\int_{0}^{\sigma}dx\,x^{2}\left(x-\sigma\right)\left[H_{0}^{(1)}\left(x\right)\mathbf{H}_{-1}\left(x\right)+H_{1}^{(1)}\left(x\right)\mathbf{H}_{0}\left(x\right)\right]\label{eq:gamma2}
\end{align}
are left to Gauss-Legendre quadrature formulas. 

When the elements $S_{m}$ and $S_{n}$ in (\ref{eq:SLdiscrete})
are adjacent, the Green function (\ref{eq:Green2D}) diverges in correspondence
of the common vertex. However, since the logarithmic divergence is
integrable and the singular end point is not considered, standard
Gauss-Legendre quadrature applies.

\subsection{Singular integrals in (\ref{eq:DLdiscrete}) and (\ref{eq:ADLdiscrete})}

As follows from the fact that $\left(\mathbf{r}-\mathbf{r}'\right)\perp\mathbf{n}$
for both $\mathbf{r}$ and $\mathbf{r}'$ lying on the same
segment with unit normal $\mathbf{n}$, all double integrals over
coincident segments in (\ref{eq:DLdiscrete}) and (\ref{eq:ADLdiscrete})
vanish identically. On the other hand, the only singular contribution
to the integrations over adjacent segments is when both $f_{i}^{m}\left(\mathbf{r}\right)$
and $f_{j}^{n}\left(\mathbf{r}'\right)$ are one at the common
vertex. In this case, the singularity is stronger than that in the
previous section and direct use of Gauss-Legendre quadrature formulas
may lead to inaccurate results. A useful technique to improve accuracy
consists in introducing a coordinate transformation with vanishing
Jacobian at the common vertex to cancel the singularity \cite{Sutradhar2008}.

\subsection{Singular integrals in (\ref{eq:HOdiscrete})}

Since a direct evaluation of the hypersingular integrals in (\ref{eq:HOdiscrete})
would require explicit cancellation of the logarithmic divergences
arising from both coincident and adjacent integrations, an alternative
approach based on the variational formulation described in \cite{Nedelec2001,Steinbach2008}
is considered in the present section. We start by writing the bilinear
form induced by the hypersingular operator (\ref{eq:HO}):
\begin{equation}
\left\langle \zeta\right|\hat{\mathsf{n}}\left|\psi\right\rangle =\int_{S}d\mathbf{r}\,\zeta\left(\mathbf{r}\right)\hat{\mathsf{n}}\left[\psi\right]\left(\mathbf{r}\right).\label{eq:bilinearForm}
\end{equation}
Then, exploiting the symmetry of the Green function and making use
of integration by parts, we have:
\begin{align}
\hat{\mathsf{n}}\left[\psi\right]\left(\mathbf{r}\right) & =\fint_{S}d\mathbf{r}'\mathrm{curl}_{S}g^{(\mathrm{2D})}\left(\mathbf{r},\mathbf{r}'\right)\mathrm{curl}_{S}'\psi\left(\mathbf{r}'\right)+k^{2}\fint_{S}d\mathbf{r}'g^{(\mathrm{2D})}\left(\mathbf{r},\mathbf{r}'\right)\psi\left(\mathbf{r}'\right)\left(\mathbf{n}\cdot\mathbf{n}'\right),
\end{align}
where:
\begin{equation}
\mathrm{curl}_{S}\equiv\mathbf{n}\cdot\mathbf{curl}=n_{x}\frac{\partial}{\partial y}-n_{y}\frac{\partial}{\partial x}\label{eq:curlS}
\end{equation}
and $\mathbf{curl}$ is the surface curl on $\mathbb{R}^{2}$. Using
again integration by parts, (\ref{eq:bilinearForm}) is reduced to:
\begin{equation}
\left\langle \zeta\right|\hat{\mathsf{n}}\left|\psi\right\rangle =\int_{S}d\mathbf{r}\fint_{S}d\mathbf{r}'g^{(\mathrm{2D})}\left(\mathbf{r},\mathbf{r}'\right)\left[k^{2}\left(\mathbf{n}\cdot\mathbf{n}'\right)\zeta\left(\mathbf{r}\right)\psi\left(\mathbf{r}'\right)-\mathrm{curl}_{S}\zeta\left(\mathbf{r}\right)\mathrm{curl}_{S}'\psi\left(\mathbf{r}'\right)\right].\label{eq:solvedBilinearForm}
\end{equation}
If we take $\zeta\left(\mathbf{r}\right)$ and $\psi\left(\mathbf{r}'\right)$
to be the basis functions $f_{i}\left(\mathbf{r}\right)$ and
$f_{j}\left(\mathbf{r}'\right)$, we can now replace (\ref{eq:HOdiscrete})
by:
\begin{align}
\mathsf{n}_{ij} & =\sum_{m\in i}\,\sum_{n\in j}\int_{S_{m}}d\mathbf{r}\fint_{S_{n}}d\mathbf{r}'g^{(\mathrm{2D})}\left(\mathbf{r},\mathbf{r}'\right)\left[k^{2}\left(\mathbf{n}\cdot\mathbf{n}'\right)f_{i}^{m}\left(\mathbf{r}\right)f_{j}^{n}\left(\mathbf{r}'\right)-\mathrm{curl}_{S_{m}}f_{i}^{m}\left(\mathbf{r}\right)\mathrm{curl}_{S_{n}}'f_{j}^{n}\left(\mathbf{r}'\right)\right].\label{eq:solvedHypersingularMatrix}
\end{align}

In order to apply the $\mathrm{curl}_{S}$ operator to (\ref{eq:basisFunctions}),
the parameterization (\ref{eq:parametricCoordinates}) is lifted into
the two-dimensional tubular neighborhood of the $n$-th segment:
\begin{equation}
\mathbf{r}_{\varepsilon}\left(t_{n}\right)=\mathbf{r}_{A}^{n}+\left(\mathbf{r}_{B}^{n}-\mathbf{r}_{A}^{n}\right)t_{n}+\varepsilon\mathbf{n}.\label{eq:parametricExtension}
\end{equation}
Then, taking the inner product of (\ref{eq:parametricExtension})
with $\left(\mathbf{r}_{B}^{n}-\mathbf{r}_{A}^{n}\right)$,
we can write:
\begin{equation}
f_{j}^{n}\left(\mathbf{r}\right)\equiv\begin{cases}
1-\frac{1}{l_{n}^{2}}\left(\mathbf{r}-\mathbf{r}_{A}^{n}\right)\cdot\left(\mathbf{r}_{B}^{n}-\mathbf{r}_{A}^{n}\right) & \mathrm{if}\:\mathbf{r}_{j}=\mathbf{r}_{A}^{n};\\
\frac{1}{l_{n}^{2}}\left(\mathbf{r}-\mathbf{r}_{A}^{n}\right)\cdot\left(\mathbf{r}_{B}^{n}-\mathbf{r}_{A}^{n}\right) & \mathrm{if}\:\mathbf{r}_{j}=\mathbf{r}_{B}^{n}.
\end{cases}\label{eq:extendedBasisFunctions}
\end{equation}
Equation (\ref{eq:extendedBasisFunctions}) provides a constant extension
of the functions (\ref{eq:basisFunctions}) along $\mathbf{n}$.
On using (\ref{eq:curlS}) and the definition of unit normal to the
$n$-th segment in $\mathbb{R}^{3}$, it follows that:
\begin{equation}
\mathrm{curl}_{S_{n}}f_{j}^{n}=\begin{cases}
-\frac{1}{l_{n}} & \mathrm{if}\:\mathbf{r}_{j}=\mathbf{r}_{A}^{n};\\
\frac{1}{l_{n}} & \mathrm{if}\:\mathbf{r}_{j}=\mathbf{r}_{B}^{n}.
\end{cases}
\end{equation}

All the integrations in (\ref{eq:solvedHypersingularMatrix}) can
now be computed as in \ref{subsec:D1}. In particular, Gauss-Legendre
quadrature formulas directly apply whenever $S_{m}\neq S_{n}$. Conversely,
the four possible integrals over coincident segments acquire the following
form:
\begin{align}
\widetilde{I}_{a}^{n} & =\widetilde{I}_{d}^{n}=k^{2}I_{a}^{n}-\frac{1}{l_{n}^{2}}\int_{S_{n}}d\mathbf{r}\fint_{S_{n}}d\mathbf{r}'g^{(\mathrm{2D})}\left(\mathbf{r},\mathbf{r}'\right)=k^{2}I_{a}^{n}-\frac{i}{2k^{2}l_{n}^{2}}\varGamma_{1}\left(kl_{n}\right);\label{eq:upsilon1122tilda}\\
\widetilde{I}_{b}^{n} & =\widetilde{I}_{c}^{n}=k^{2}I_{b}^{n}+\frac{1}{l_{n}^{2}}\int_{S_{n}}d\mathbf{r}\fint_{S_{n}}d\mathbf{r}'g^{(\mathrm{2D})}\left(\mathbf{r},\mathbf{r}'\right)=k^{2}I_{b}^{n}+\frac{i}{2k^{2}l_{n}^{2}}\varGamma_{1}\left(kl_{n}\right),\label{eq:upsilon1221tilda}
\end{align}
where $I_{a}^{n}$, $I_{b}^{n}$ and $\varGamma_{1}\left(kl_{n}\right)$
are defined in (\ref{eq:I11I22})-(\ref{eq:gamma0}) and the tilde
is to avoid notation overlap.

\section{Calderon preconditioning of (\ref{eq:operatorHdiscrete})\label{sec:E}}

The boundary integral operators (\ref{eq:SL})-(\ref{eq:HO}) can
be shown to satisfy the Calderon relations \cite{Nedelec2001}:
\begin{align}
\hat{\mathsf{d}}\,\hat{\mathsf{s}} & =\hat{\mathsf{s}}\,\hat{\mathsf{d}}^{\dagger};\label{eq:Calderon1}\\
\hat{\mathsf{n}}\,\hat{\mathsf{d}} & =\hat{\mathsf{d}}^{\dagger}\hat{\mathsf{n}};\\
\hat{\mathsf{d}}^{2}-\hat{\mathsf{s}}\,\hat{\mathsf{n}} & =\hat{I}/4;\\
\hat{\mathsf{d}}^{\dagger2}-\hat{\mathsf{n}}\,\hat{\mathsf{s}} & =\hat{I}/4,\label{eq:Calderon4}
\end{align}
with $\hat{I}$ representing the identity operator. After discretizing the
boundary $S$, both sides of (\ref{eq:SL})-(\ref{eq:HO}) can be
expanded on a set of node-based basis functions $\left\{ f_{j}\right\} $.
Let us consider, for instance, equation (\ref{eq:SL}):
\begin{equation}
\hat{\mathsf{s}}\left[f\right]\left(\mathbf{r}_{S}\right)=\sum_{j}\alpha_{j}\,f_{j}\left(\mathbf{r}_{S}\right);\quad f\left(\mathbf{r}_{S}'\right)=\sum_{j}\beta_{j}\,f_{j}\left(\mathbf{r}_{S}'\right).
\end{equation}
Applying the Galerkin method, we obtain:
\begin{equation}
\sum_{j}F_{ij}\,\alpha_{j}=\sum_{j}\mathsf{s}_{ij}\,\beta_{j}
\end{equation}
being $F$ the Gram matrix defined in (\ref{eq:F}) and $\mathsf{s}$
the discrete operator in (\ref{eq:SLdiscrete}). The same procedure
can be used to discretize the other operators and their products,
leading to the following matrix version of (\ref{eq:Calderon1})-(\ref{eq:Calderon4}):
\begin{equation}
\mathbf{F}^{-1}\mathbf{h}\,\mathbf{F}^{-1}\mathbf{h}=\mathbf{I}/4,\label{eq:discreteCalderon}
\end{equation}
where:
\begin{equation}
\mathbf{h}_{ij}\equiv\left(\begin{array}{cc}
-\mathsf{d}_{ij} & \mathsf{s}_{ij}\\
-\mathsf{n}_{ij} & \mathsf{d}_{ij}^{\dagger}
\end{array}\right);\quad\mathbf{F}_{ij}\equiv\left(\begin{array}{cc}
F_{ij} & 0\\
0 & F_{ij}
\end{array}\right).
\end{equation}
Using (\ref{eq:discreteCalderon}), (\ref{eq:singleLayerExpression})-(\ref{eq:hypersingularExpression})
and (\ref{eq:operatorHdiscrete}), it can be shown that \cite{Niino2012}:
\begin{equation}
\mathbf{H}\,\mathbf{F}^{-1}\mathbf{H}\,\mathbf{F}^{-1}\approx\mathbf{I}+\mathbf{C},
\end{equation}
with $\mathbf{C}$ resulting from the discretization of a compact
operator. Therefore, the block matrix $\mathbf{F}$ can be employed
as a right preconditioner in both (\ref{eq:scattMatrixEquDiscrete})
and (\ref{eq:matrixEquSpectralDiscrete}).\\\\\\


\providecommand{\newblock}{}

\end{document}